\documentclass[prb,aps,longbibliography,twocolumn,floatfix]{revtex4-2}
\usepackage{hyperref}
\usepackage{subfigure}
\usepackage{graphicx}
\usepackage{verbatim}
\usepackage{multirow}
\usepackage{amsmath}
\newcommand{\beq}{\begin{eqnarray}}
\newcommand{\eeq}{\end{eqnarray}}
\usepackage[usenames, dvipsnames]{color}
\DeclareMathOperator{\Tr}{Tr}
\def\blue{\color{blue}}

\usepackage{physics}	
\usepackage{epstopdf}
\usepackage{bbold}
\def \bs{\textbf}

\begin{document}

\title{
Bogoliubov Fermi Surfaces  in Spin-1/2 Systems: Model Hamiltonians and Experimental Consequences}
\author{Chandan Setty$^1$}
\thanks{email for correspondence: csetty@ufl.edu}
\author{Yifu Cao$^1$}
\author{Andreas Kreisel$^2$}
\author{Shinibali Bhattacharyya$^1$}
\author{P. J. Hirschfeld$^1$
}

\begin{abstract}
\begin{center}
$^1$Department of Physics, University of Florida, Gainesville, Florida, USA

$^2$Institut f\"ur Theoretische Physik Universit\"at Leipzig
	D-04103 Leipzig, Germany
	\end{center}
	\vskip .3cm
Bogoliubov Fermi surfaces (BFSs) are topologically protected regions of zero energy excitations in a superconductor whose dimension equals that of the underlying normal state Fermi surface.  Examples of Hamiltonians exhibiting this ``ultranodal'' phase are known to preserve charge-conjugation ($C$) and parity ($P$) but break time-reversal ($T$).  In this work, we provide examples of model Hamiltonians that do not necessarily preserve this symmetry pattern but have well-defined sign-changing Pfaffians yielding BFSs. While their topological character has not been recognized previously, some of the models we present have been extensively studied in prior literature. We further examine thermodynamic and electronic properties arising from the ultranodal state.  In particular, we study the effect of a weak Zeeman field close to the topological transition and propose distinguishing features of BFSs using residual specific heat and tunneling conductance. Our calculation of the superfluid density in a toy multi-band model indicates a window of interband pairing strength where BFSs are stable with a positive superfluid density.  We also present additional signatures of BFSs in spin-polarized spectral weight and total magnetization measurements.

\end{abstract}

\maketitle
\section{Introduction} 
The dimensionality of zero-energy quasiparticle excitations forms a defining characteristic of superconductors (SCs) both from the perspective of their pairing symmetry as well as their experimental phenomenology. Most well-known conventional or unconventional SCs are either fully gapped, or can host line- or point-nodes that have dimensionality  strictly less than that of the underlying Fermi surface (FS). However, in the presence  of certain combinations of discrete symmetries, SCs can acquire extended nodes called Bogoliubov Fermi Surfaces (BFSs) -- defined as contours of zero-energy quasiparticle excitations in the Brillouin zone that share dimensionality with the normal state FS.  The notion of a BFS, while not new~\cite{Sondhi1998,Kivelson2008}, has witnessed a recent resurgence~\cite{Timm2017, Brydon2017, Brydon2018, Setty2019} in multiband systems due to the recognition of its topological protection -- the zero-energy quasiparticle excitations are robust to the introduction of a finite intra-pocket pair (independent of its symmetry). This must be contrasted with the topologically trivial case where zero-energy quasiparticles stem from pure inter-pocket pairs far away from the FS \footnote{These states are energetically unfavorable compared to those with gaps formed at the FS}. In the latter, the quasiparticle spectrum is gapped out by even an infinitesimally small and isotropic intra-pocket order parameter.  \par
A formal description of BFSs~\cite{Timm2017,Brydon2018} hinges on the existence of a $\mathbb{Z}_2$ topological invariant which can be expressed as a Pfaffian, Pf$(\bf k)$. The Pfaffian is well-defined and acquires purely real values if the Hamiltonian can be similarity-transformed into a basis where it is antisymmetric. If the Pfaffian changes sign at any point(s) in the Brillouin zone, a BFS is guaranteed.  As argued by the authors of Ref. \onlinecite{Timm2017}, a sufficient condition (but not necessary) for the existence of such a transformation is a Hamiltonian with charge-conjugation ($C$) and parity ($P$). In addition, a time reversal symmetry breaking (TRSB) pairing component in spin space (or type-2 TRSB) can lead to a sign change in Pf$(\bf k)$ and a BFS emerges. For spin-$\frac{1}{2}$ multi-band systems (in the absence of external fields), when the intra-band pairing exceeds a critical value set by the strength of the inter-band pair, the BFS is destroyed \cite{Setty2019}.\par
Possible material realization of BFSs was first suggested~\cite{Timm2017} in the context of higher spin angular momentum systems with $j=3/2$ pairing and additional intrinsic TRSB. These include uranium (URu$_2$Si$_2$, UPt$_3$) and strontium (SrPtAs, Sr$_2$RuO$_4$)  based compounds which have multiple bands crossing the Fermi level. Very recent proposals also include spin liquids~\cite{Yang2020} and superfluid $^3$He~\cite{Eltsov2020}. The spin $s=1/2$ iron chalcogenide superconductor FeSe$_{1-x}$S$_x$ was argued to be the first system to exhibit a topological transition into an ``ultranodal'' phase with BFSs in the absence of external fields~\cite{Setty2019}. The large spin-orbit coupling (SOC) in the parent FeSe~\cite{Borisenko2016} can in principle stabilize a spin-triplet inter- and intra-band pairing~\cite{VafekChubukov, EugenioVafek, CvetkovicVafek2013, Kee2012}, and with the additional presence of non-unitary TRSB, the sign change condition of the Pfaffian is satisfied as the inter-pocket pair exceeds intra-pocket pair at some momenta. 
 Empirically, the intra-pocket gaps become more nodal as a function of sulfur doping and the resulting zeros of the Pfaffian occur close to the original normal state FS. However, despite experimental evidence of TRSB order parameter in many of the aforementioned compounds including FeSe~\cite{Matsuura2019APS} and additional work being done on the sulfur doped FeSe~\cite{Fujiwara2019APS}, so far there is no direct attempt to probe its non-unitary character in any of them. \textcolor{black}{While theoretical proposals have also been made in Weyl systems~\cite{Sigrist2017,Herbut2020}, including experimental signatures in dichalcogenides~\cite{Kanigel2020}, more candidate models and materials are needed to fully explore the physics of BFSs in greater depth}. One of the main goals of this paper is to explore experimental methods of identifying the ultranodal state definitively.  \par
 
The most obvious observable manifestation of the topological phase is  a large enhancement of tunneling density of states (DOS) at zero energy as well as large $T=0$ residual specific heat $(C_V/T)$ and thermal conductivity ($\kappa$) even in very clean samples~\cite{Matsuda2018,Hanaguri2018,Shibauchi2020,FeSe_review2020}.  As a practical matter, however, it can be difficult to distinguish the ultranodal state from a nodal system with disorder by measures of residual density of states alone, unless one is certain disorder effects are very weak.
 Further contrasts between the topological phase and nodal SCs are expected to show up in the low energy/temperature behavior of the single-particle tunneling rate, magnetic penetration depth and NMR spin-relaxation rate. The power-law temperature and frequency dependences are succinctly summarized in a recent work~\cite{Timm2019}.  Nevertheless, the most direct evidence for BFSs  can be obtained from a combination of angle-resolved photoemission (ARPES) measurements and  probes of non-unitary TRSB.  While the former has been difficult due to low temperatures and poor sample surfaces, the latter, as we argue below, may be hard to detect due to relatively low values of internal TRSB moments.    \par
 We begin our work by studying model Hamiltonians that do not necessarily preserve charge-conjugation or parity but  have well defined Pfaffians that can change sign nevertheless. This helps extend the phase space of material systems that can also exhibit the physics of BFS. Some of the models are simple and well-known in the existing literature, but we choose to revisit them in order to highlight their topological characteristics. We then examine electronic and thermodynamic properties of BFSs including their spectral properties, effect of in-plane and out-of-plane Zeeman field close to the topological transition, magnetization, and BFS stability via electromagnetic response and band structure effects. We perform these calculations assuming no residual interactions between quasiparticles~\cite{Moon2019}. We find that, close to the topological transition, an in-plane field always  pushes the system into the topological phase irrespective of its direction. The thermodynamic properties, however, depend on the directionality of an out-of-plane field -- the system becomes topological or trivial depending on whether the applied field aligns or cancels the internal TRSB field. We present results for the DOS and residual specific heat for each of the cases above. While it has been argued that known models that support BFSs can be unstable due to their negative superfluid density~\cite{QHWang2009}, our calculation of the electromagnetic response shows that the superfluid density can remain positive even in the presence of a BFS provided the inter-pocket pairing is below a critical value
 set by aspects of the band structure such as the band masses.  A further increase in inter-pocket pairing strength then indeed renders the ultranodal state unstable due to a negative superfluid density.  Furthermore, the total internal magnetization of the toy model studied in~ Ref. \onlinecite{Setty2019} shows that the non-unitary TRSB magnetic moment is relatively small and could be non-trivial to detect experimentally.  Finally, we examine the stability of BFSs due to changes in the electronic structure  and find that small energy band separations are more susceptible to formations of  BFSs. \par
 \textcolor{black}{In Section II we outline the four generic model Hamiltonians along with their corresponding symmetries, Pfaffians and conditions for existence of BFSs. In Section III, we focus specifically on the low-energy effective Hamiltonian in Ref.~\cite{Setty2019} used to describe the physics of Fe(Se,S), and expand the study of various experimental consequences. Section IV analyses electronic structure effects which can determine the formation of BFSs in addition to the Pfaffian. We conclude with our final remarks in Section V.} \par
 
 \section{Model Hamiltonians}\label{ModelHamiltonians}
 In this section we review simple models, some previously studied in the literature, that have extended FSs in the superconducting state associated with their $\mathbb{Z}_2$ topological invariant. In each of the cases, we show that their Pfaffians are real and well-defined, and analyze the conditions under which they can change sign. We consider four models: two of them with explicit TRSB in the kinetic energy and two other with TRSB in the pairing terms. \par
 \textit{a) d-wave SC in Zeeman field:} The first model we consider is that of a one-band $d$-wave superconductor in an external Zeeman field~\cite{Sondhi1998}. We begin with the total Hamiltonian $\hat{H} = { \frac 12} \sum_{\bs k} \Psi_{\bs k}^{\dagger} H(\bs k) \Psi_{\bs k}$, where the Nambu operator is defined in the basis $\Psi_{\bs k}^{\dagger} = \bigl(c_{\bs k \uparrow}^{\dagger},c_{\bs k \downarrow}^{\dagger}, c_{-\bs k \uparrow}, c_{-\bs k \downarrow} \bigr)$ and $c_{\bs k \sigma}^{\dagger}$ is the electron creation operator with momentum $\bs k$ and spin $\sigma$. The individual terms in the Hamiltonian are expanded as $\hat{H} = \hat{H}_0 +\hat{H}_{\mathrm Z}^j+ \hat{H}_{\Delta}$.  We choose the normal state part of the Hamiltonian as $\hat{H}_0 = \sum_{\bs k\sigma} \epsilon(\bs k) c_{\bs k \sigma}^{\dagger} c_{\bs k  \sigma} $ written in the band basis with $\epsilon({\bs k})=k^2/2m$, $\hat{H}_{\mathrm Z}^{j} =\sum_{\bs k\sigma\bar \sigma}h_j\sigma^{\sigma\bar\sigma}_j ~ c_{\bs k \sigma}^{\dagger} c_{\bs k  \bar \sigma}  $ is the Zeeman term with a constant magnetic field $h_j$ along direction $j=x,y,z$, and $\hat{H}_{\Delta}$ is a spin-singlet pairing Hamiltonian $\hat H_\Delta = \sum_{\bs k}\Delta({\bs k})c_{\bs k\uparrow}^\dagger c_{-\bs{k}\downarrow}^\dagger$ +\textit{h.c}. with $d$-wave order parameter $\Delta(\bs k) = \Delta_0 \cos( 2\phi_{\bs k})$. In the basis of $\Psi_{\bs k}^{\dagger}$ and a magnetic field along the $z$ direction ($h_z \equiv h$), the total Hamiltonian takes the form
 \beq \nonumber
  H(\bs k) &=& \Delta(\bs k) \left(i\pi_y \otimes i\sigma_y\right) + \epsilon(\bs k) (\pi_z \otimes \sigma_0) +h (\pi_z \otimes \sigma_z)\\
 &&
 \label{Zeeman}
\eeq
where $\sigma_i$ and $\pi_i$ are Pauli matrices in spin and particle-hole space. For a constant $h$ independent of momentum, the Hamiltonian above maintains $C$ and $P$ symmetries individually. The Pfaffian of the Hamiltonian is real and well-defined, and given by $\text{Pf}(\bs k) = \epsilon(\bs k)^2 + \Delta(\bs k)^2 - h^2$. As a functional of the band structure, the Pfaffian acquires arbitrarily large positive values, but has a minimum given by $\text{Min}\{{\text{Pf}(\bs k)}\} = \Delta(\bs k)^2 - h^2$. Hence for a nodal SC, the Pfaffian changes sign for an infinitesimally small Zeeman field near the nodal points giving rise to BFSs, consisting of nodal loops circling the $h=0$ point node in 2D. \par
 \textit{b) Loop currents coexisting with d-wave order:} From the analysis above, it can be seen that in order for the Pfaffian to change sign, we do not need to require the momentum dependence of the field $h$ to have any particular symmetry with respect to inversion.  As an example of such a state studied in literature, we consider the loop current order coexisting with $d$-wave superconductivity which was proposed as a possible superconducting ground state of the underdoped cuprates~\cite{Varma1997, Varma1999}. In the presence of intra-plaquette loop currents, the hopping parameters pick up additional phases originating from the flux and the total Hamiltonian takes a form similar to Eq.~(\ref{Zeeman}) but with a momentum dependent effective magnetic field. Such a Hamiltonian can be written as 
  \begin{align} \nonumber
 H(\bs k) =& \Delta(\bs k) \left(i\pi_y \otimes i\sigma_y\right) + \epsilon(\bs k) (\pi_z \otimes \sigma_0) \nonumber\\
 &+ J(\bs k)(\pi_z \otimes \sigma_z),
 \label{Loop}
\end{align}
where $J(\bs k) = - J(-\bs k)$ is odd under inversion and is proportional to the loop current order parameter $J$. For a square lattice, the functional form is given by $J(\bs k ) = J\left( \sin k_x - \sin k_y + \sin(k_y - k_x)\right)$. Due to this property of $J(\bs k)$ appearing as a diagonal element, the loop current term breaks both $C$ and $P$ symmetries but maintains the product $CP$. Nonetheless, stable BFSs exist as the Pfaffian minimum $\text{Min}\{{\text{Pf}(\bs k)}\} = \Delta(\bs k)^2 - J(\bs k)^2$ changes sign close to the $d$-wave nodes. \par
\textit{c) Type-1 TRSB, odd-parity, spin-triplet pair terms:} As argued in Ref.~\onlinecite{Setty2019}, for spin $\frac{1}{2}$ particles with even parity intra-and inter-pocket pairing along with a non-unitary TRSB component, a BFS is ensured if the inter-pocket pairing exceeds a critical value. Below we show that a similar argument holds when the intra-pocket pair is an odd-parity, spin-triplet, provided it is purely imaginary (type-1 TRSB). Together with real inter-pocket even parity pairs, these terms make the total pairing Hamiltonian have a mixed parity. We choose a total Hamiltonian as $\hat{H} = \hat{H}_0  + \hat{H}_{\Delta} ={\frac 12} \sum_{\bs k} \Psi_{\bs k}^{\dagger} \left(H_0(\bs k) +  H_{\Delta}(\bs k) \right)\Psi_{\bs k}$, with the normal state part written in band basis given by $\hat{H}_0= \sum_{i\sigma\bs k} \epsilon_i(\bs k) c_{\bs k i\sigma}^{\dagger} c_{\bs k i \sigma} $, where $i=1,2$ is the pocket index. For the pairing, we choose two different pairing Hamiltonians 
\beq \nonumber
{\underbar {$H$}}_{\Delta_1}(\bs k) &=& i\gamma(\bs k) (\tau_0 \otimes \sigma_x )+\Delta_0 (i\tau_y \otimes \sigma_0) +\delta (i\tau_y \otimes \sigma_z),\\\nonumber
\underbar{$H$}_{\Delta_2}(\bs k) &=& i\gamma(\bs k) (\tau_0 \otimes \sigma_x )+ \Delta_0 (\tau_x \otimes i \sigma_y) +\delta (i\tau_y \otimes \sigma_x),\\
&&
\label{Case3}
\eeq
where $\tau_i$ are the Pauli matrices in band/pocket space and underlined quantities represent one block in particle-hole space. Here ${\underbar {\textit H}}_{\Delta_1}(\bs k)$ $\left({\underbar {\textit H}}_{\Delta_2}(\bs k)\right)$ contains inter-pocket terms, with coefficients $\Delta_0$ and $\delta$, that are TRS or TRSB spin-triplet pairs with equal (opposite) spin. The term proportional to $\gamma(\bs k)$ in each case is an intra-pocket spin-triplet pair that is odd under inversion, i.e., $\gamma(\bs k) = -\gamma(-\bs k)$, and real.
Hence both pairing Hamiltonians in Eq.~(\ref{Case3}) break parity and maintain charge-conjugation, but their Pfaffians are well-defined and real. Minimizing the Pfaffian functional for the total Hamiltonian with respect to the band energies, we obtain for the two cases above  
\beq \nonumber
\text{Min}\{\text{Pf}(\bs k)\}_{\Delta_1} &=& 4 \delta^2 (\gamma(\bs k)^2 - \Delta_0^2) \\ \nonumber
\text{Min}\{\text{Pf}(\bs k)\}_{\Delta_2} &=& 4 \Delta_0^2 (\gamma(\bs k)^2 - \delta^2).
\nonumber
\eeq 
Hence BFSs exist above a critical value of ($\Delta_0$, $\delta$) set by the intra-pocket triplet pair $\gamma(\bs k)$. \par
\textit{d) Broken inversion symmetry:} In all the previous examples considered above, the individual pairing terms and basis functions for the superconducting order parameters were eigenstates of parity operator with eigenvalues $\pm 1$ (although the total pairing Hamiltonian was allowed to be mixed under parity). As a final example, we consider the scenario where inversion symmetry is explicitly broken {by the inter-pocket pair} so that this no longer holds.  The total Hamiltonian is again chosen as $\hat{H} = \hat{H}_0  + \hat{H}_{\Delta} ={\frac 12} \sum_{\bs k} \Psi_{\bs k}^{\dagger} \left(H_0(\bs k) +  H_{\Delta}(\bs k) \right)\Psi_{\bs k}$, with the normal state part written in band basis as $\hat{H}_0 = \sum_{i\sigma} \epsilon_i(\bs k) c_{\bs k i\sigma}^{\dagger} c_{\bs k i \sigma} $, and the off-diagonal pairing block in Pauli matrix notation as
\beq \nonumber
\underbar{$H$}_{\Delta}(\bs k) &=& \phi(\bs k) (\tau_0 \otimes i\sigma_y )+ \Delta_0 (\tau_x \otimes i \sigma_y) +\delta (i\tau_y \otimes i\sigma_y),\\ 
&&
\label{Case4}
\eeq
where $\phi(\bs k)$ is an even function of $\bs k$. The first term is an ordinary spin-singlet and the second term has a matrix structure similar to the inter-pocket pairing term appearing in Eq.~(\ref{Case3}). However, and in contrast to Eq.~(\ref{Case3}), the presence of the $i\tau_y\otimes i\sigma_y$ matrix in the $\delta$ term ensures that inversion symmetry is explicitly broken 
\textcolor{black}{since a state exhibiting a combination of $\Delta_0$ and $\delta$ terms will no longer be an eigenstate of the parity operator.}
Similar to the previous cases, the Pfaffian is real with an arbitrarily large and positive maximum value. The minimum, on the other hand, can be evaluated for $\delta \neq 0$ as $\text{Min}\{\text{Pf}(\bs k)\} = 4 \delta^2 \left(\phi(\bs k)^2 - \Delta_0^2\right)$, hence yielding BFSs at the appropriate sign change regions in the Brillouin zone. \textcolor{black}{A summary of the symmetries in the four models discussed above is shown in Table~\ref{SummaryTable}.}
\begin{table}[tb]
\centering
\textcolor{black}{
 \begin{tabular}{|l | c c c|} 
 \hline
 Case & C & P & T \\ [1ex] 
 \hline\hline
 a) $d$ wave + Zeeman & 1 & 1 & 0 \\ 
 \hline
 b) $d$ wave + Loop currents & 0 & 0 & 0 \\
 \hline
 c) Type 1 TRSB (triplet) & 1 & 0 & 0 \\
 \hline
 d) Mixed Parity & 0 & 0 & 0 \\ [1ex] 
 \hline
\end{tabular}
\caption{Table of summary of symmetries in the four models.}
\label{SummaryTable}
}
\end{table}

 \section{Experimental Consequences}

 \vskip .2cm
 
  In this section we study properties of BFSs that can be manifested in experiments. \textcolor{black}{To be more specific, from now on we focus on the Hamiltonian that has been proposed in the context of the Fe(Se,S) system~\cite{Setty2019}, where a model of hole and electron pockets at the $\Gamma$ and $X,Y$ points has been used to describe the electronic structure of this iron-based material}. The normal state Hamiltonian is $\hat{H}_0 = \sum_{\bs k i\sigma} \epsilon_i(\bs k) c_{\bs k i\sigma}^{\dagger} c_{\bs k i \sigma} $
where $\epsilon_i(\bs k)$ are parabolic bands centered at $\Gamma$ and $X,Y$.
  The superconducting order parameter consists of a momentum-dependent intra-band pairing term $\Delta_j(\bs k) = \Delta_{ja}(\bs k) + \Delta_j$ which is parametrized by isotropic $\Delta_j$ and an anisotropic term of the form $\Delta_{ja}({\bs k})= \Delta_{ja}(k_x^2-k_y^2)$. As a function of sulfur doping, the isotropic component of the order parameter becomes smaller in magnitude as observed experimentally (see~ Ref. \onlinecite{Setty2019} and references therein). \textcolor{black}{Additionally, we introduce the BFS pairing ansatz through a time-reversal preserving triplet component $\Delta_0$ and a time-reversal breaking triplet component $\delta$. For simplicity, we set them equal in for the rest of the discussion, $\Delta_0=\delta$. }
\begin{figure*}[ht!] 
    \centering
    \subfigure[]{\includegraphics[width=0.253\textwidth]{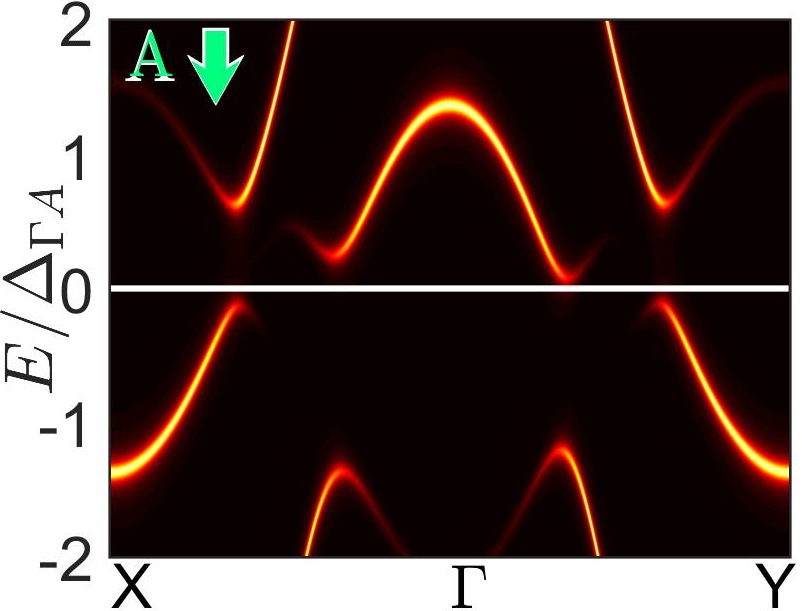}\label{AD}} \,
    \subfigure[]{\includegraphics[width=0.22\textwidth]{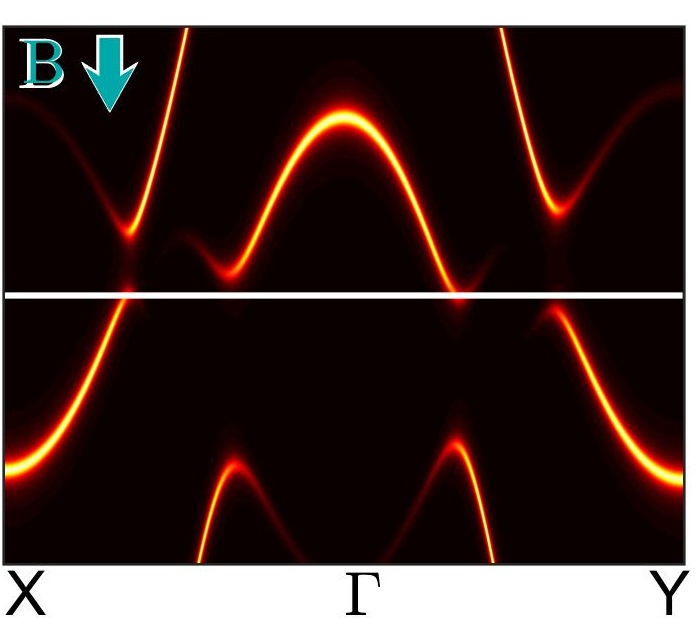}\label{BD}} \,
    \subfigure[]{\includegraphics[width=0.22\textwidth]{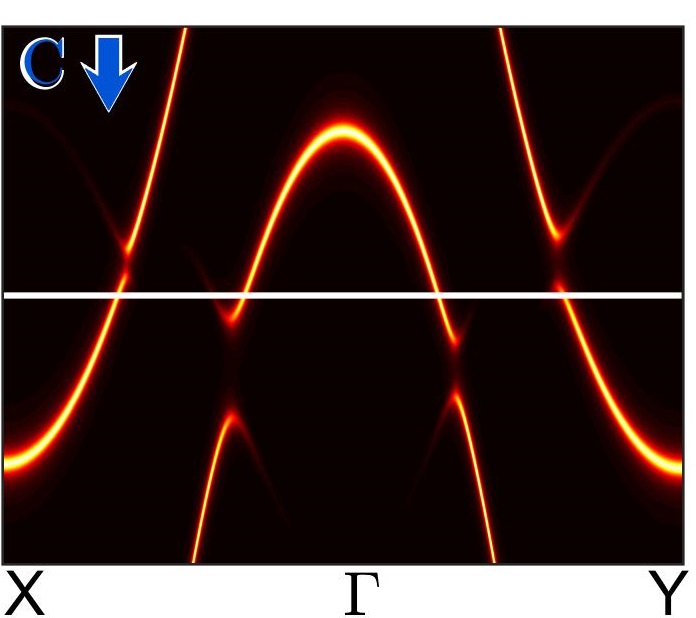}\label{CD}} \,
    \subfigure[]{\includegraphics[width=0.247\textwidth]{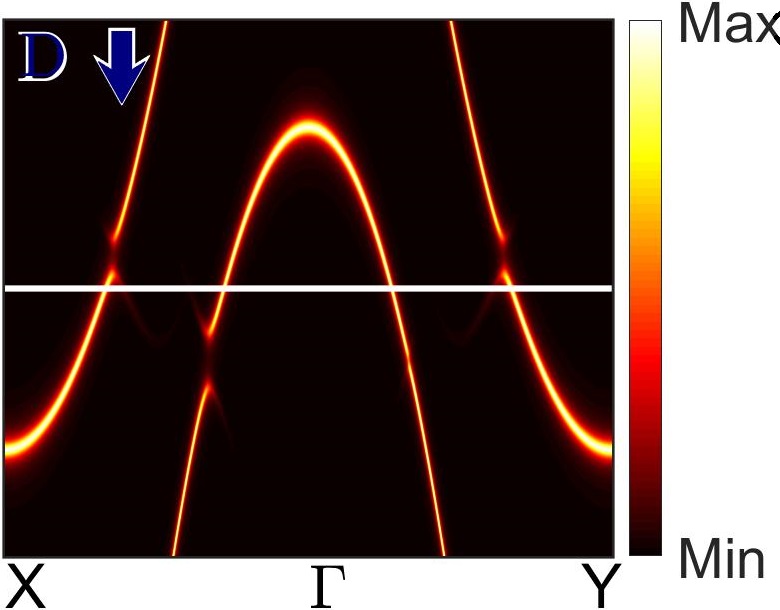}\label{DD}} \,
    \subfigure[]{\includegraphics[width=0.253\textwidth]{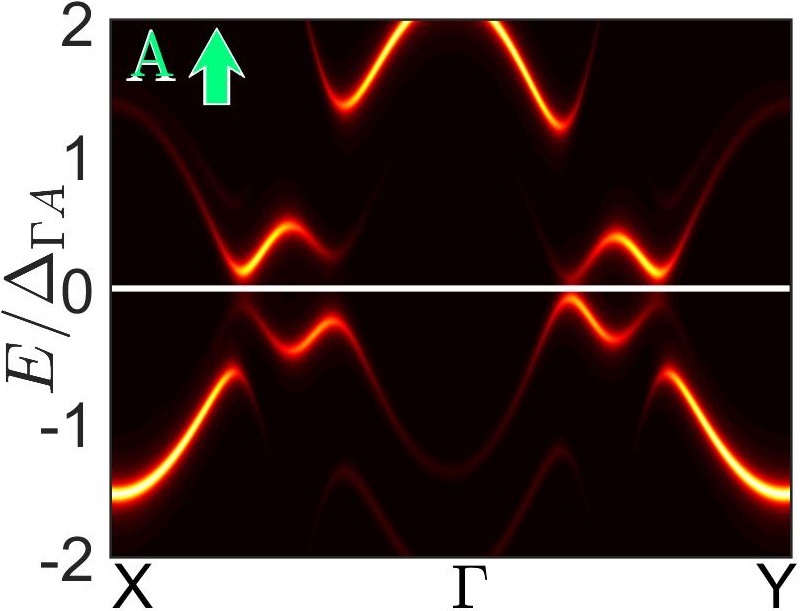}\label{AU}} \,
    \subfigure[]{\includegraphics[width=0.22\textwidth]{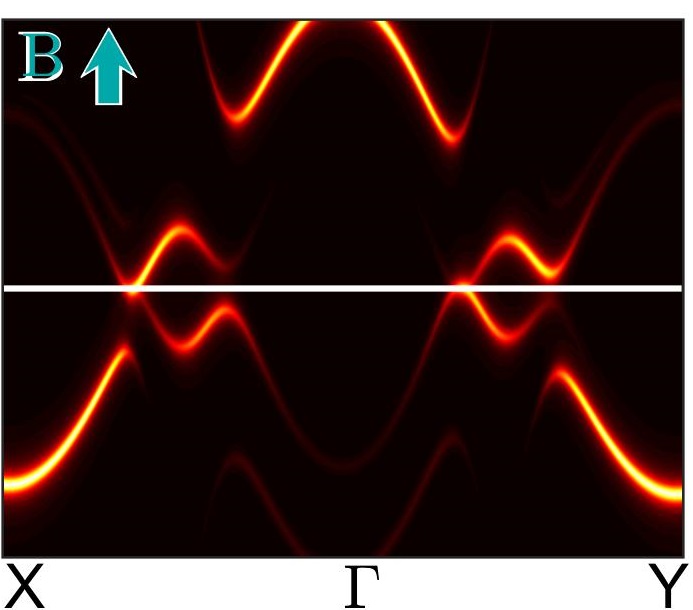}\label{BU}} \,
    \subfigure[]{\includegraphics[width=0.22\textwidth]{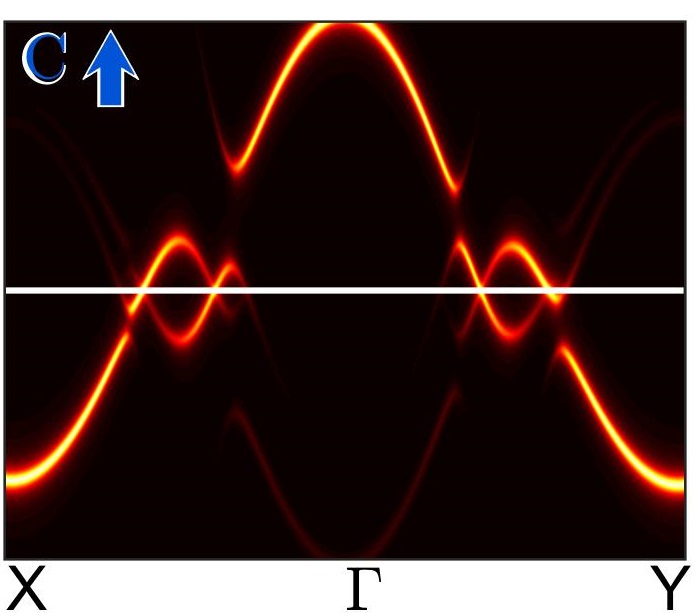}\label{CU}} \,
    \subfigure[]{\includegraphics[width=0.247\textwidth]{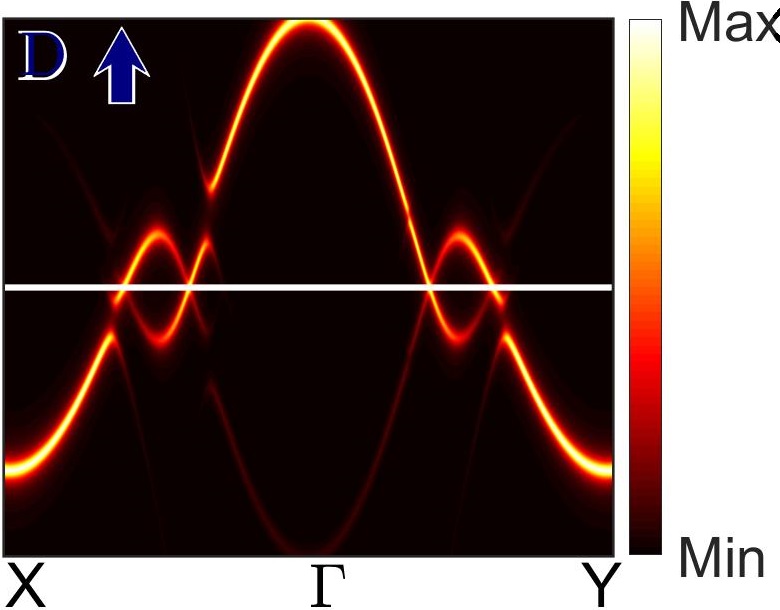}\label{DU}} \,
	\caption{Spin-resolved spectral function $A_\sigma(\mathbf{k}, \omega)$ for the ``realistic'' model described in the text, appropriate for Fe(Se,S), evaluated along high-symmetry path $X-\Gamma-Y$ at temperature $T/T_{\mathrm{c}}=0.02$ for the example cases\cite{Setty2019} (a) A, (b) B, (c) C, and (d)D (with the energy axis normalized to hole pocket anisotropic gap maximum $\Delta_{\Gamma A}$) (see text for parameters corresponding to different cases). The arrow pointing downwards refers to spin-down component $\sigma= \downarrow$.  (e-h) Same as in (a-d) but for spin-up component $\sigma= \uparrow$.}
    \label{fig_ABCD_Arpes}
\end{figure*}
The pairing term in the band basis then reads
  \beq \nonumber
\hat{H}_{\Delta} &=& { \frac 12} \Delta_0\sum_{i< j, \bs k} \left(c_{\bs ki\uparrow}^{\dagger}c_{-\bs kj\uparrow}^{\dagger}+ c_{\bs ki\downarrow}^{\dagger}c_{-\bs kj\downarrow}^{\dagger}\right) + h.c. - (i \leftrightarrow j)\\ \nonumber
&&+{ \frac 12} \delta  \sum_{i < j, \bs k} \left(c_{\bs ki\uparrow}^{\dagger}c_{-\bs kj\uparrow}^{\dagger}-c_{\bs ki\downarrow}^{\dagger}c_{-\bs kj\downarrow}^{\dagger}\right) + h.c. - (i \leftrightarrow j)\\ 
&& + {\frac 12} \sum_{i, \bs k} \Delta_i(\bs k) \left(c_{\bs ki\uparrow}^{\dagger}c_{-\bs ki\downarrow}^{\dagger}-c_{\bs ki\downarrow}^{\dagger}c_{-\bs ki\uparrow}^{\dagger}\right) + h.c.\,, 
\label{pairing_H}
\eeq
where the additional indices on the fermionic operators label bands. For the calculations to follow, we use the generic Hamiltonian stated above unless stated otherwise.
\subsection{Spectral functions}
\label{sec_spectral}
With a workable model in place, we begin by calculating the spin polarized spectral functions that can be measured by ARPES. The following calculations are performed by taking simple parabolic dispersion for the electronic structure. \textcolor{black}{Each of the pockets are chosen to have a quadratic dispersion with momenta defined in units of the  inverse lattice constant},
\begin{subequations}
\begin{align}
 \epsilon_\Gamma(\bs k) &= -\frac{4\alpha}{\pi^2} \bs k^2 +E_+  \label{eq_dispersion_two_pocket_1}\\
\epsilon_X(\bs k)&= \frac{4\alpha}{\pi^2}\Bigl[\bigl (\frac{k_x-\pi}{1+\epsilon}\bigr)^2+ \bigl(\frac{k_y}{1-\epsilon})^2\Bigr] -E_-\\
 \epsilon_Y(\bs k)&= \frac{4\alpha}{\pi^2}\Bigl[ \bigl(\frac{ k_x}{1-\epsilon}\bigr)^2+\bigl(\frac{k_y-\pi}{1+\epsilon}\bigr)^2\Bigr] -E_-
 \label{eq_dispersion_two_pocket_3}
\end{align}
\end{subequations}
with the parameters $\alpha=2$ and $E_+=0.6$ , $E_-=0.6$ $\varepsilon=0.2$ and additionally inserting symmetry related electron bands having band minima at $(0,-\pi)$ and $(-\pi,0)$. \textcolor{black}{Here and below, energies are given in arbitrary units; where applicable, we plot quantities that are associated with the dimension of energy normalized to the hole pocket anisotropic gap maximum $\Delta_{\Gamma A}=0.5$, which is associated with the coherence peak of 2.2 meV\cite{Sprau2017,Hanaguri2018}.} The order parameters are explicitly given by
\begin{subequations}
\begin{align}
\Delta_\Gamma(\bs k)&=\Delta_\Gamma +\frac{4\Delta_{\Gamma a}}{\pi^2}(k_x^2-k_y^2)\\
\Delta_X(\bs k)&=\Delta_X +\frac{4\Delta_{Xa}}{\pi^2}\Bigl[-\bigl(\frac{k_x-\pi}{1+\epsilon}\bigr)^2+\bigl(\frac{k_y}{1-\epsilon}\bigr)^2\Bigr]\\
\Delta_Y(\bs k)&=\Delta_Y +\frac{4\Delta_{Ya}}{\pi^2}\Bigl[\bigl(\frac{ k_x}{1-\epsilon}\bigr)^2-\bigl(\frac{k_y-\pi}{1+\epsilon}\bigr)^2\Bigr]
\end{align} 
\end{subequations}
and the corresponding order parameters on the symmetry related electron bands. We use for the anisotropic gap components $\Delta_{\Gamma a} = 0.1$ , and $\Delta_{X a} = \Delta_{Y a} = 0.4$. The isotropic gap components are denoted as [$\Delta_{\Gamma},\Delta_{X},\Delta_{Y} $] and are assumed to decrease continuously as a function of sulfur doping, and we choose the same values as in Ref.\onlinecite{Setty2019}, i.e. we define sets of parameters A-D with
A:[$0.40,0.35,0.35$],
B:[$0.35,0.27,0.35$],
C:[$0.16,0.20,0.25$],
D:[$0.07,0.07,0.07$].
 These parameters were adopted to describe a situation where the gap in Fe(Se,S) evolves from a highly anisotropic, nematic state with nodes along one axis of each Fermi surface pocket, to  an even more anisotropic state with four nodes on the $\Gamma$ pocket as the tetragonal phase is reached, consistent with experiment\cite{Matsuda2018,Hanaguri2018,Shibauchi2020,FeSe_review2020}.

We begin by evaluating the spin dependent intensities as measured by ARPES. Diagonalizing the Hamiltonian, we obtain the eigenenergies $E_\mu(\mathbf{ k})$ of the Bogoliubov quasiparticles in $\mu^{\mathrm{th}}$ band and a unitary transformation with the matrix elements $a_\mu^{j\sigma}(\mathbf{k})$ such that the spin-resolved spectral function reads
\begin{align}
 \begin{split}
    A_\sigma(\mathbf{k}, \omega) & = - \frac{1}{ \pi }  \text{Tr Im} \left(  G^\sigma_{11}(\mathbf{k},\omega) \right)  \\
    &= \frac{1}{ \pi } \sum_{\mu}  \frac{ \eta \, | a_\mu^{1\sigma}(\mathbf{k}) |^2}{\eta^2 + (\omega - E_\mu(\mathbf{k}))^2 }.
 \end{split}
\end{align}
where $G^\sigma_{11}(\mathbf{k},\omega)$ refers to the diagonal Gorkov Green's function,  $\sigma$ is the spin index and $\eta $ is an artificial broadening parameter. 
In Fig.\ref{fig_ABCD_Arpes}, we show the spin-resolved spectral function along the high-symmetry path $X-\Gamma-Y$ for down-spin component and up-spin components for cases A-D. Case A (Fig.\ref{AD} and \ref{AU}) shows that the system will be fully gapped without any residual BFSs. With evolution in sulfur doping content of the system, as mimicked in the transition from case A to D, the spectral map reveals a Fermi level crossing of the bands as can be seen in Fig.\ref{BD}-\ref{DD}, \ref{BU}-\ref{DU}. The BFSs become larger as more momentum space points satisfy the Pfaffian sign change condition. Such features can be easily detected in ARPES measurements. With spin-resolved ARPES, it is possible to probe into different momentum sections of the same band as depicted in Fig.\ref{fig_ABCD_Arpes}. 
The calculation of spectral function was carried out on a momentum path of size $1200$ points in each segment of $X-\Gamma$ and $\Gamma-Y$, and on a frequency grid of $1500$ points. The artificial broadening $\eta$ was set to $0.008$. Further analysis of the effect of magnetic field on the spectral function is presented in Section \ref{sec_zeeman}.

\subsection{Spontaneous Magnetization}
 \begin{figure}[tb]
\includegraphics[width=\linewidth]{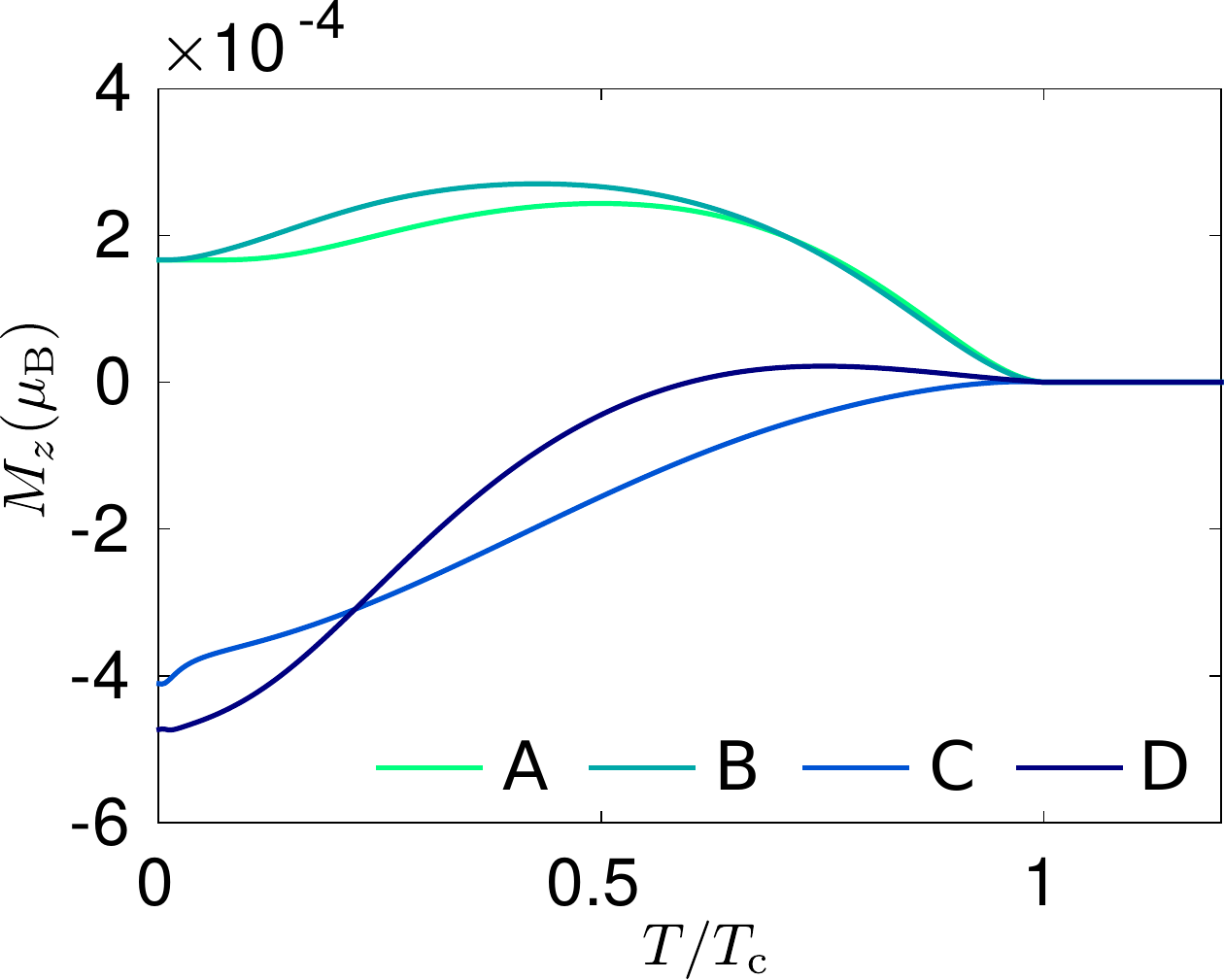}
\caption{Magnetization in $z$ direction from the time reversal symmetry breaking pairing term. Assuming a BCS like behavior of all order parameters in Eq. (\ref{pairing_H}), one obtains a small but in general nonzero magnetization in the superconducting state. The sign of the magnetization and its functional behavior depends on details of the band structure and the order parameters as seen for the example cases A-D\cite{Setty2019}.}
\label{fig_mag_field}
\end{figure}
 In this sub-section we calculate the expectation value $\langle \mathbf{ M}\rangle$ of the sum of the spin operators $\mathbf{S}_{\mathbf{ k}}={\sum_{i,\alpha,\beta}} \sigma_i^{\alpha \beta}\mathbf{ e}_i c_{\mathbf{ k}\alpha}^\dagger c_{\mathbf{ k}\beta} $  in the superconducting state
\begin{equation}
 \mathbf{M}=\frac{1}{N_{k}}\sum_{\mathbf{ k}}\mathbf{ S}_{\mathbf{ k}}\,.
\end{equation}
The expectation values $\langle c_{\mathbf{ k}\alpha}^\dagger c_{\mathbf{ k}\beta}\rangle$ at finite temperature are evaluated in the basis where the Hamiltonian is diagonal,
\begin{equation}
 \langle c_{\mathbf{ k}\alpha}^\dagger c_{\mathbf{ k}\beta}\rangle = \sum_{\mu}
 a_\mu^{1\alpha}(\mathbf{k})^*
  a_\mu^{1\beta}(\mathbf{k}) n(E_\mu(\mathbf{ k})),
\end{equation}
 such that the matrix elements
$a_\mu^{j\sigma}(\mathbf{k})$ from the unitary transformation and the Fermi function $n(x)=1/(\exp(x/T)+1)$  of the eigenenergies $E_\mu(\mathbf{ k})$ enter.
In the following, we use the same gap parameter values as in the previous section for cases A-D. For our model system proposed for Fe(Se,S)~\cite{Note}, it turns out that there is only a $z$ component of the magnetization due to the choice of the time reversal symmetry breaking. Indeed, at $T>T_{\mathrm c}$ the spontaneous magnetization vanishes, and acquires a finite value once $T<T_{\mathrm c}$. Note that the details of the magnetization curves depend on the electronic structure, i.e. the relative size, position of holelike bands and electronlike bands and their densities of states as well as the balance of interband pairing and intraband pairing contributions. As it can be seen in Fig. \ref{fig_mag_field}, for the different choices of the intraband pairing A-D, the behavior of the magnetization and also its value at $T\rightarrow 0$ can be different.
 The agreement of the magnetization at $T\rightarrow 0$ for cases A and B is accidental for our choice of parameters and the negative value for the cases C and D is due to the additional contribution from BFS stemming from the electron bands which in our choice have a larger density of states and a dominant negative contribution to $M_z$ if BFS are present.  To summarize, there is a finite magnetization as expected from the TRSB pairing. However, the direction and temperature dependence of this quantity is not directly connected to the topological state with BFS. \textcolor{black}{We find that the value of the magnetization can be small, especially if contributions from  hole and electron bands compensate partially (see Fig.~\ref{fig_mag_field}).
Therefore, the screening due to the Meissner effect might not allow the direct detection of the magnetization at all if the corresponding field is smaller than $H_{c1}$.
 }

\subsection{Zeeman field}
\label{sec_zeeman}
 
\begin{figure*}[ht!] 
    \centering
    \subfigure[]{\includegraphics[width=0.265\textwidth]{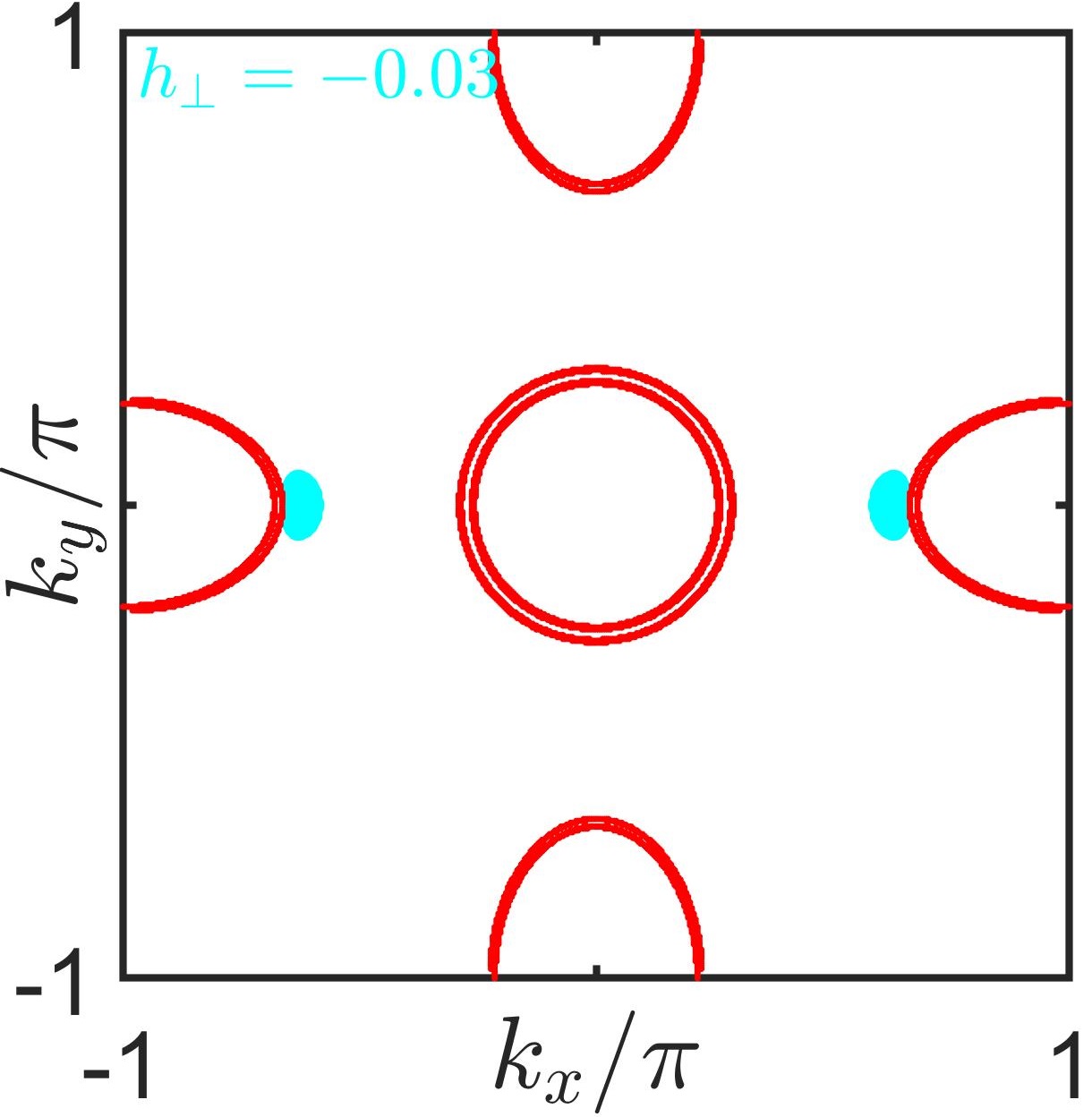}\label{FS1}} \;
    \subfigure[]{\includegraphics[width=0.35\textwidth]{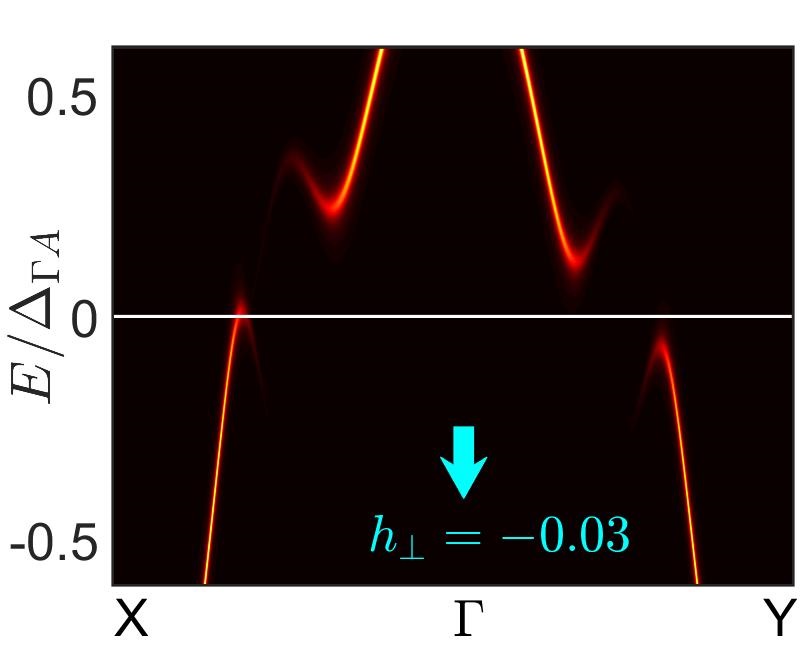}\label{ZAD1}}
    \subfigure[]{\includegraphics[width=0.35\textwidth]{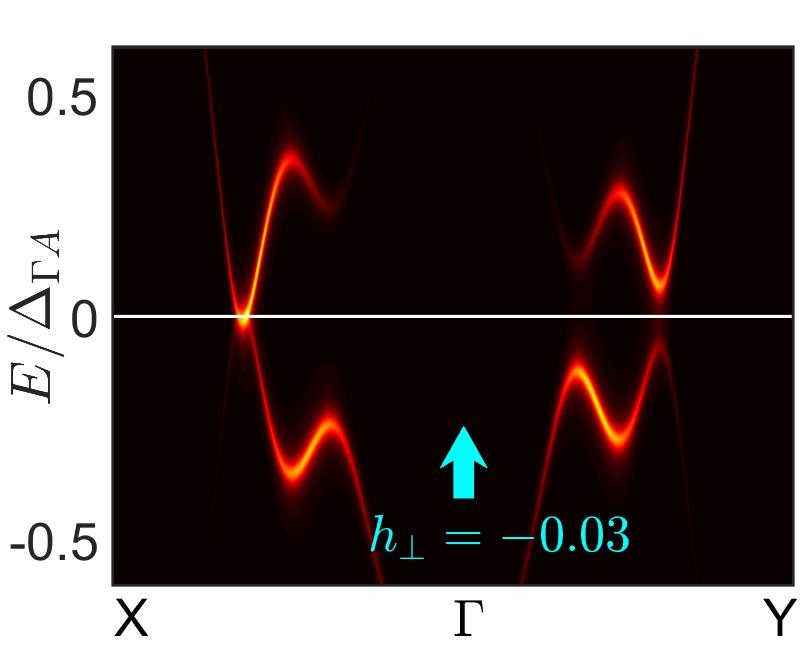}\label{ZAU1}}
    \subfigure[]{\includegraphics[width=0.265\textwidth]{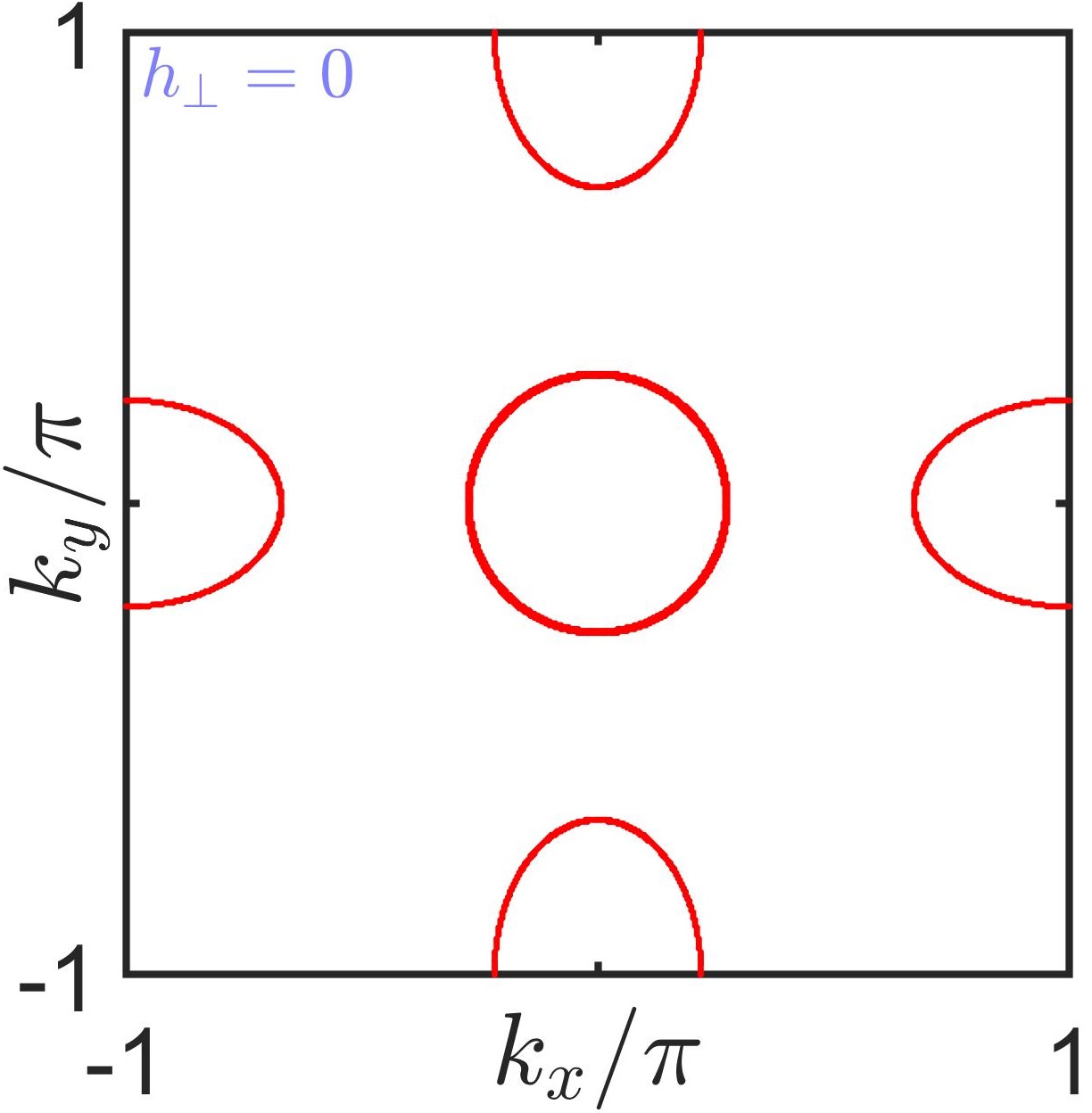}\label{FS2}} \;
    \subfigure[]{\includegraphics[width=0.35\textwidth]{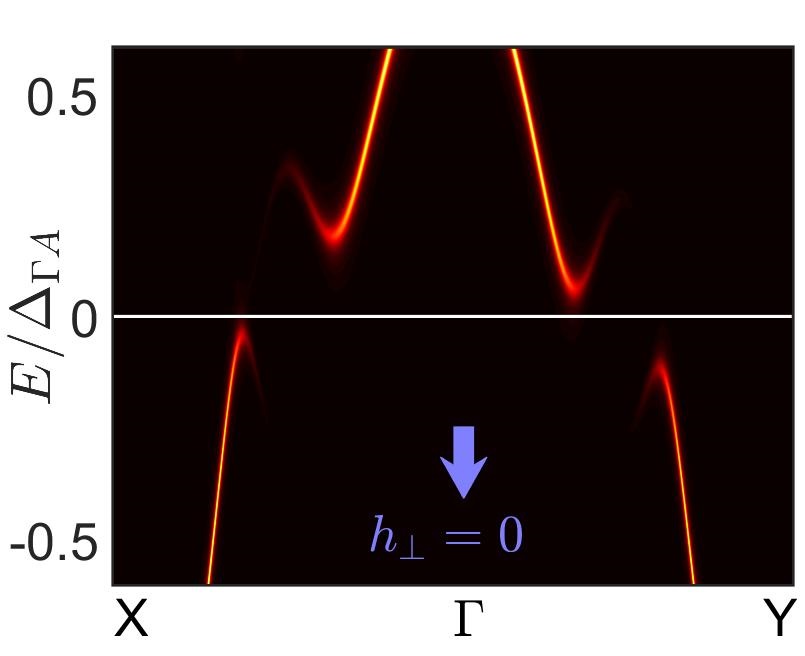}\label{ZAD2}}
    \subfigure[]{\includegraphics[width=0.35\textwidth]{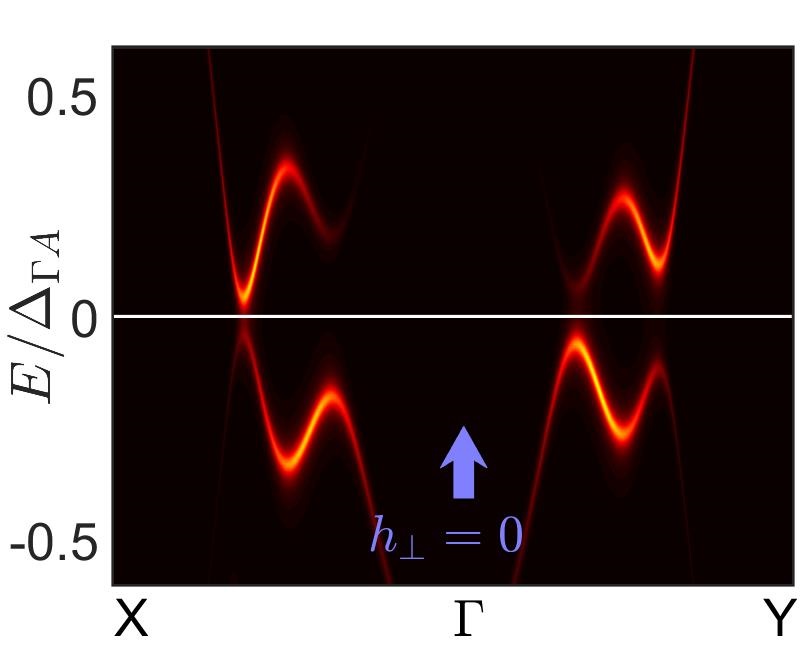}\label{ZAU2}}
    \subfigure[]{\includegraphics[width=0.265\textwidth]{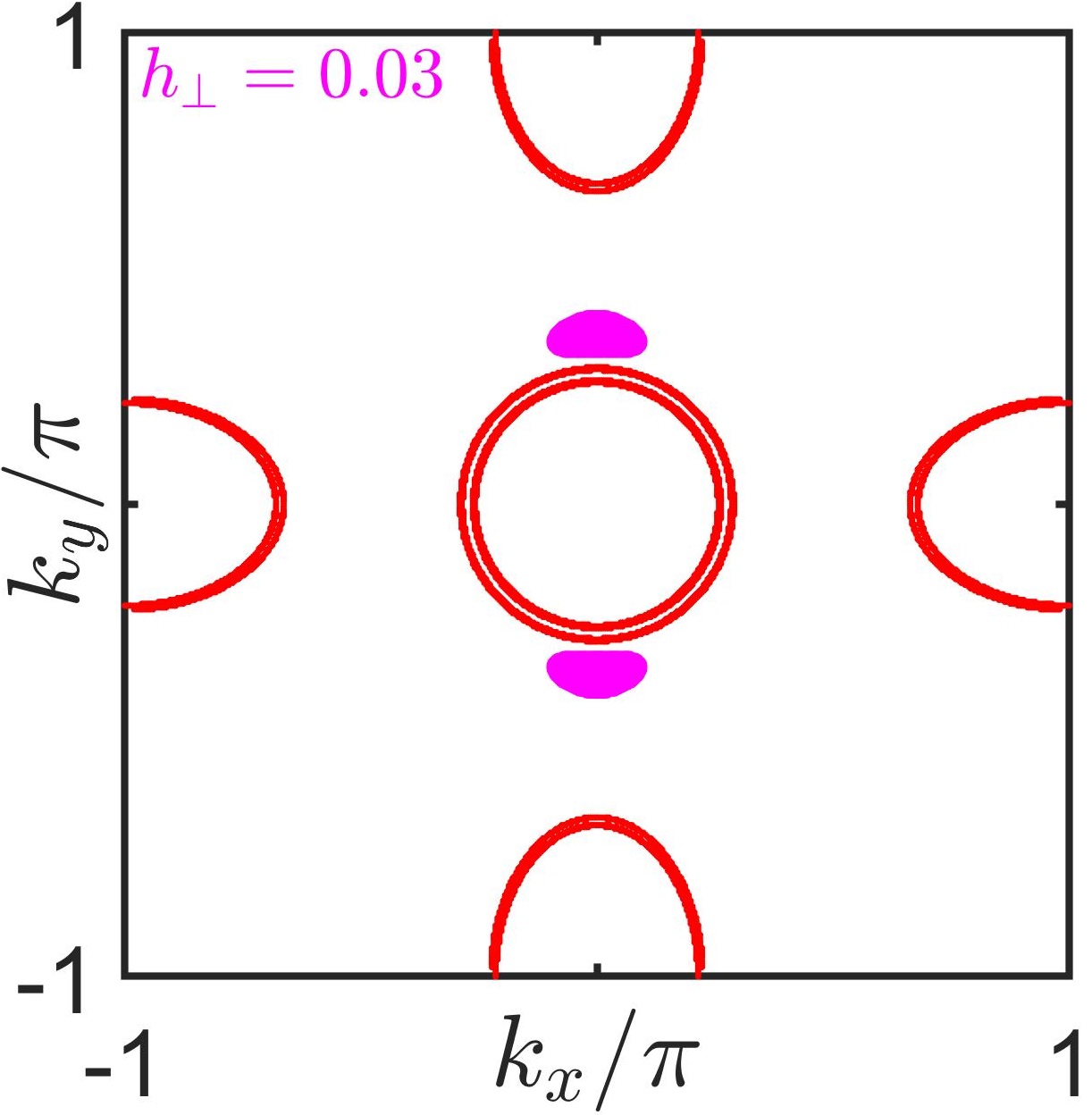}\label{FS3}} \;
    \subfigure[]{\includegraphics[width=0.35\textwidth]{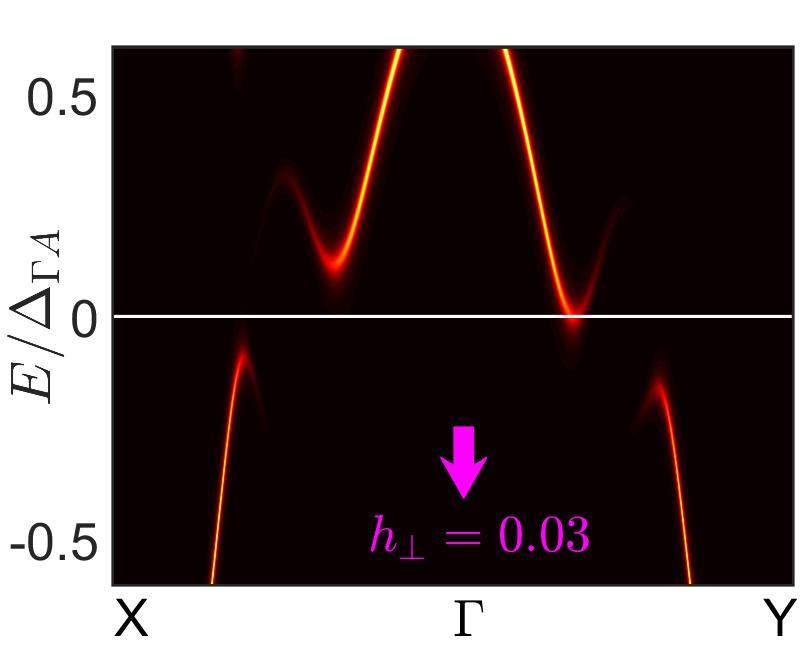}\label{ZAD3}}
    \subfigure[]{\includegraphics[width=0.35\textwidth]{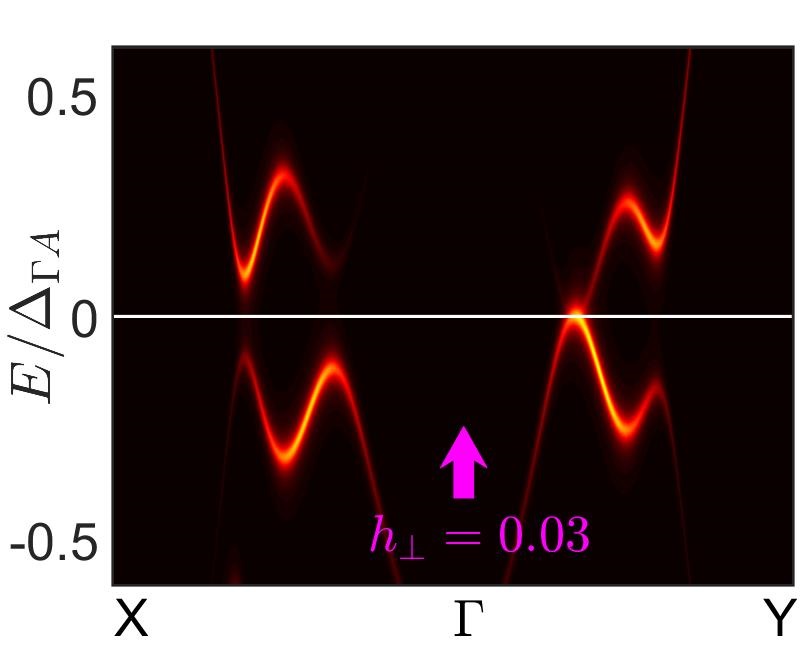}\label{ZAU3}}
	\caption{(a) Normal state Fermi surface (red contour) and Bogoliubov Fermi surface in superconducting state (blue/pink patches) for magnetic field $h_\perp = -0.03$,
	(b) Spin-resolved spectral function $A_\sigma(\mathbf{k}, \omega)$ evaluated along high-symmetry path $X-\Gamma-Y$ at temperature $T/T_{\mathrm{c}}=0.02$ (with the energy axis normalized to hole pocket anisotropic gap maximum $\Delta_{\Gamma A}$). The arrow pointing downwards refers to spin-down component $\sigma= \downarrow$.  (c) Same as in (b) but for spin-up component $\sigma= \uparrow$.
	(d-f) Same as in (a-c) but for magnetic field $h_\perp = 0$. (g-i) Same as in (a-c) but for magnetic field $h_\perp = +0.03$. Note that while results are plotted over a putative 1st Brillouin zone, the model is actually continuous. Note that the sign of the magnetic field $(\pm h_\perp)$ is chosen with respect to the sign of the inter-band gap component $\Delta_0$.}
    \label{fig_FS_ARPES}
\end{figure*}

We now study the effect of a weak Zeeman field on BFSs close to the topological transition. Before we present results for our more realistic distribution of bands specific to the iron superconductor Fe(Se,S), we consider a simple toy model which includes the Zeeman field to demonstrate the underlying physics. We choose this model to be of the form
 \beq
  \hat{H} &=& \hat{H}_0  + \hat{H}_{\Delta} + \hat{H}^{j}_{\mathrm Z} \\ \nonumber
   &=&{ \frac 12} \sum_{\bs k} \Psi_{\bs k}^{\dagger} \left(H_0(\bs k) +  H_{\Delta}(\bs k) + H_{\mathrm Z}^j(\bs k)\right)\Psi_{\bs k}.
  \eeq
Here $j=x, y, z$ is the direction of the magnetic field
and the Zeeman term reads explicitly $\hat{H}_{\mathrm Z}^{j} =\sum_{\bs k\sigma\bar \sigma}h_j\sigma^{\sigma\bar\sigma}_j ~ c_{\bs k \sigma}^{\dagger} c_{\bs k  \bar \sigma}  $.
The normal state part written in band basis given by $\hat{H}_0= \sum_{i\sigma\bs k} \epsilon_i(\bs k) c_{\bs k i\sigma}^{\dagger} c_{\bs k i \sigma} $. For the pairing, we work with two special cases -- sign-change $(+-)$ and no-sign-change $(++)$ pairing on the two pockets. These pairing terms are written as
\beq \nonumber
{\underbar {$H$}}_{\Delta}^{++}(\bs k) &=& \Delta(\bs k) (\tau_0 \otimes i\sigma_y )+\Delta_0 (i\tau_y \otimes \sigma_0) +\delta (i\tau_y \otimes \sigma_z),\\ \nonumber
{\underbar {$H$}}_{\Delta}^{+-}(\bs k) &=& \Delta(\bs k) (\tau_z \otimes i\sigma_y )+\Delta_0 (i\tau_y \otimes \sigma_0) +\delta (i\tau_y \otimes \sigma_z),\\
&&
\eeq
which take a form similar to ${\underbar {$H$}}_{\Delta_1}(\bs k)$ written above in Section~\ref{ModelHamiltonians} c), but with a real $\Delta(\bs k)$ and $\Delta(\bs k) = \Delta(-\bs k)$. The $\Delta(\bs k)$ term defines the pairing amplitude for a spin-singlet intra-pocket pair with the same order parameter magnitudes on both pockets.  We choose this pairing form for purposes of illustration but our conclusions can be easily extended to the case of the pairing Hamiltonian ${\underbar {$H$}}_{\Delta_2}(\bs k)$ (See Eq.~\ref{Case3}) with a similar spin-singlet intra-pocket pair. Finally, the Zeeman term is written in the expanded particle-hole basis as $H_{\mathrm Z}^{j}(\bs k) = h_j \left(\pi_z \otimes \tau_0 \otimes \sigma_j \right)$. When the magnetic field is in-plane ($h_{\parallel}$, $j=x,y$), the relevant Pfaffian minima with respect to the band energies are
\begin{align}
\mathop{\mathrm{Min}}\{\mathop{\mathrm{Pf(\bs k)}}\} =\begin{cases}
    \text{$4\delta^2\left(\Delta(\bs k)^2 - \Delta_0^2 - h_{\parallel}^2\right)$}~~~~~~~~~\text{ $++$ }\\
       \text{$4\Delta_0^2\Delta(\bs k)^2 - 4 \delta^2 \left(\Delta_0^2 + h_{\parallel}^2\right)$}~~~~\text{ $+-$.} 
     \end{cases} \label{Minimum}
\end{align}
Hence an in-plane field always pushes the system into the topologically non-trivial state for both phase distributions as long as $\delta \neq 0$. Moreover, and as one should anticipate, this conclusion is independent of the direction of in-plane field. For the case when the field is out-of-plane ($h_{\perp}$, $j=z$), the total Pfaffian is written as a sum of two terms -- one quadratic in the field and another linear. It takes the form
\beq \nonumber
\text{Pf(\bs k)}_{\pm}\{\delta, \Delta(\bs k), \Delta_0, h_{\perp}, \epsilon_i \}& =& \text{Pf(\bs k)}_{2,\pm}\{\delta, \Delta(\bs k), \Delta_0, h_{\perp}^2, \epsilon_i \}  \\
&&- h_{\perp} \delta \Delta_0\left( \epsilon_1 + \epsilon_2\right)
\label{eqn_pfaffian}
\eeq
where the first term, $\text{Pf(\bs k)}_{2,\pm}$,  is quadratic in the field and $\pm$ denotes the sign-change and no-sign-change cases respectively. For a given set of bands with dispersion $\epsilon_i$, the relative signs of the field $h_{\perp}$ and the TRSB component $\delta$ determines the sign of the linear term. While for generic field strengths both terms are important in determining the existence of BFSs, the linear term dominates the physics at small fields. In this limit, one can ignore the field dependence of $\text{Pf(\bs k)}_{2,\pm}$ and we obtain
\beq \nonumber
\text{Pf(\bs k)}_{\pm}\{\delta, \Delta(\bs k), \Delta_0, h_{\perp}, \epsilon_i \}& \simeq & \text{Pf(\bs k)}_{2,\pm}\{\delta, \Delta(\bs k), \Delta_0, 0, \epsilon_i \}  \\
&&- h_{\perp} \delta \Delta_0\left( \epsilon_1 + \epsilon_2\right).
\label{PfaffianTCP}
\eeq
Close to the topological critical point, we know that the first term $\text{Pf(\bs k)}_{2,\pm}\{\delta, \Delta(\bs k), \Delta_0, 0, \epsilon_i \} \simeq 0$. Hence whether the Pfaffian changes sign in this regime is completely determined by the linear term in $h_{\perp}$, i.e., the relative signs of $h_{\perp}$ and $\delta$. If the Pfaffian has a certain sign for a given direction of the weak field, it \textit{must} change sign when the direction of the field is flipped. Therefore, if there exists no BFS for a certain direction of the field, $h_{\perp}$, one must emerge for $-h_{\perp}$. This statement is independent of the details of the band structure $\epsilon_i$, provided one has purely electron or hole like pockets and the field has little effect on the internal electronic structure of the material, and therefore, forms a distinct signature of topological phase transition. The above signatures are expected to show up in the specific heat and tunneling DOS close to the topological critical point.\par
In the case where one has mixed electron and hole pockets as is the case with FeSe, the situation is less unambiguous. As is evident from the last term in Eq.~\ref{PfaffianTCP}, the relative sign between the two pockets with respect to the Fermi level becomes important. In such a scenario, the electron and hole pockets satisfy the Pfaffian sign change condition separately for \textit{opposite} direction of the field. Hence, close to the topological transition, BFSs form only on the hole pockets for one direction of the field and on the electron pockets for the opposite direction. Nonetheless, a key of signature of BFSs would manifest in the asymmetry of the residual specific heat and tunneling conductance spectra with respect to flipping the field direction as the hole and electron pockets generally have different density of states at the Fermi energy (see the next sub-section on specific heat and residual differential conductance).

We now present the spectral map of the system evaluated under the presence of a magnetic field  close to the topological transition. Choosing the same model as discussed in Section \ref{sec_spectral}, we select the isotropic gap components as $\Delta_\Gamma = 0.23, \Delta_X = 0.28, \Delta_Y = 0.33$, inter-band gap component to be $\Delta_0 = 0.3$ and time-reversal broken component $\delta = \Delta_0$. Anisotropic gap components are $\Delta_{\Gamma a} = 0.1$ , and $\Delta_{X a} = \Delta_{Y a} = 0.3$. This choice of parameters is made such that the system is very close to the transition into the topological state. Exactly at the transition, the BFS reemerges upon application of an infinitesimal magnetic field as seen in Eq.\ref{eqn_pfaffian}. We use three values of magnetic field $h_\perp$ to generate BFS as seen in Fig.\ref{FS1} $h_\perp = -0.03$, \ref{FS2} $h_\perp = 0$ and \ref{FS3} $h_\perp = +0.03$. Note that the sign of the magnetic field $(\pm h_\perp)$ is chosen with respect to the sign of the inter-band gap component $\Delta_0$. The red lines denote the normal-state FS contour under the same parameter values for magnetic field. Notice that the spin-degeneracy is lifted under the presence of the magnetic field, giving rise to very closely-spaced concentric Fermi pockets in the normal state. Case \ref{FS2} shows no BFS under zero magnetic field close to the topological transition. We recover ultranodal BFS states on the electron pockets in case \ref{FS1} (blue) with negative magnetic field and on the hole pockets in case \ref{FS3} (pink) with positive field. 

The corresponding spectral function for $h_\perp = 0$ case (Fig.\ref{ZAD2} and \ref{ZAU2}) shows that the system will be fully-gapped without any residual Bogoliubov surfaces. With the application of $- h_\perp$, the residual BFS appears along $\Gamma-X$ direction which is shown in the Fermi level crossing of the spectral map in Fig.\ref{ZAD1} and \ref{ZAU1}. Upon flipping the direction of the magnetic field to $+ h_\perp$, the Fermi level crossing shifts towards the $\Gamma-Y$ direction as shown in Fig.\ref{ZAD3} and \ref{ZAU3}. 
 Since the effects described here are small but observable, it is worth stating clearly that the best chance for a ``smoking gun" experiment where the BFS is induced by an external probe requires operation very close to the transition point.   

\subsection{Specific heat and differential conductance}

To obtain specific heat, one can start from calculating the entropy $S$ of the free Fermi gas of Bogoliubov quasiparticles, in terms of the eigenenergies $E_\mu (\bs k)$ of the Hamiltonian $ H(\bs k)$ and use $C_V=T^{-1} dS/dT$ to obtain
\begin{align}
\begin{split}
C_V = 
\frac{2}{N_k} \sum_{\mu, \mathbf k} & \frac{n(E_\mu(\mathbf{k}))n(-E_\mu(\mathbf{k})) }{T^2} \\
&\times \left( E_\mu(\mathbf{k})^2 -T E_\mu(\mathbf{k}) \frac{\partial E_\mu(\mathbf{k}) }{\partial T} \right)
\end{split}   
\end{align}
where $\mu$ is a band index and \mbox{$n(x)=1/(\exp(x/T)+1)$} is the Fermi function. The temperature dependence of the order parameter is assumed to follow a mean field behavior {\color{black} with dimensionless function $d(T)=\tanh(1.76\sqrt{T_c/T-1})$, where we have set $T_c=0.15$  which is chosen to yield $2\Delta_{\Gamma A}/T_c \approx 7$ i.e. larger than the BCS value and very close to the ratio discussed for FeSe\cite{Shibauchi2020}}.
 \begin{figure}[tb]
\includegraphics[width=\linewidth]{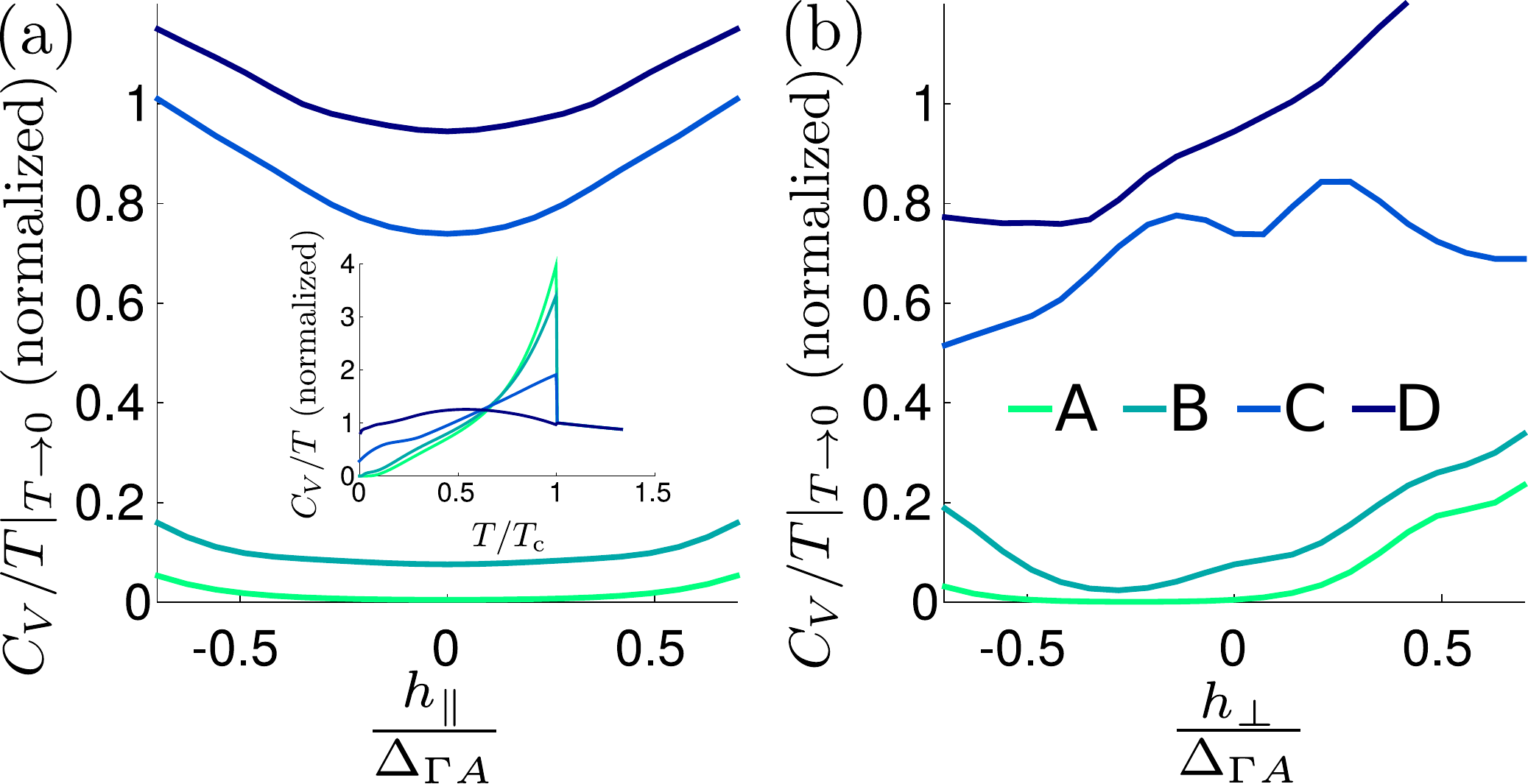}
\caption{ Residual specific heat $C_V/T$ at zero temperature, normalized to the value in the normal state. For fields in plane, $h_\parallel$ the $C_V/T$ is an even function of the field (a), while for out of plane field, $h_\perp$, there are also contributions odd in the field (b). The explicit behavior depends on the details of the order parameter (and the band structure); here we present results for a model for Fe(Se,S)\cite{Setty2019} with parameters for the cases A-D as given in the text. The inset shows the expected zero field behavior as function of temperature as also presented in Ref. \cite{Setty2019}.
} \label{CvbyTpic}
\end{figure}
{\color{black}
Results of $\left .(C_V/T)\right|_{T\rightarrow0}$ as a function of fields in plane $h_\parallel$ and out of plane $h_\perp$ are shown in Fig.\ref{CvbyTpic}. The numerical evaluations are carried out at a low temperature of $T/T_{\mathrm c}=0.0007$.
While for in plane fields, the specific heat is an even function of the field (see Fig.\ref{CvbyTpic} (a)), it also acquires a dependence on odd powers of the field $h_\perp$ in Fig.\ref{CvbyTpic} (b); finite values of $\left .(C_V/T)\right|_{T\rightarrow0}$ are signatures of the ultranodal state. Starting from the case $A$, where no Bogoliubov Fermi surface exists, it is possible to tune into the topological state by fields in any direction, while the field in $z$ direction is more effective. In the parameter set $B$ it is not possible to decrease  $\left .(C_V/T)\right|_{T\rightarrow0}$ with fields in plane; in contrast the curve for $h_\perp$ has a finite slope at zero field such that leaving the topological state might be possible. Note that qualitatively similar behavior of $C_V/T$ is expected at any temperature $T\leq T_c$; this quantity will be an even function of the in plane fields, but acquire odd powers of $h_\perp$.
 We have not calculated the corrections to the specific heat due to the low energy states in the vortex phase of the superconductor (e.g. Volovik contribution from extended  states\cite{Volovik1993}, or Caroli-de Gennes-Matricon states in the core\cite{CdGM1964}). However, we stress that these will always increase the value of $C_V/T$ and are independent  of the direction of the field; they  therefore do not change the conclusion that in the state with BFS, $C_V/T$ acquires dependence on $h_\perp$ of odd powers.
A momentum grid of  $1200 \times 1200$ was used to calculate the specific heat to yield convergence at the lowest temperature.}

Also the density of states $\rho(E)$ and thus the STM conductance $dI(V)/dV$ exhibits similar
symmetry behavior as function of the magnetic field in plane $h_\parallel$ and out of plane $h_\perp$.
To highlight this behavior, we calculate the differential conductance $dI(V)/dV$ as a function of external bias voltage $V$ and external field. For this, we first calculate the density of states by
\begin{align}
\rho(E)  &= \frac{1}{N_k} \sum_{{\bf k},\sigma} A_\sigma({\bf k},E)
\end{align}
and then perform a convolution with the derivative of the Fermi function to obtain the differential conductance
\begin{align}
 \frac{dI(V)}{dV} & \propto \int_{-E_L}^{E_U} \frac{ \rho(E) e^{(E-eV)/T} }{ \left( 1 + e^{ (E-eV)/T} \right)^2 } dE \,,
\end{align}
where the solution of $\rho(E)$ is used to evaluate $\frac{dI(V)}{dV}$.
\begin{figure}[tb]
\includegraphics[width=\linewidth]{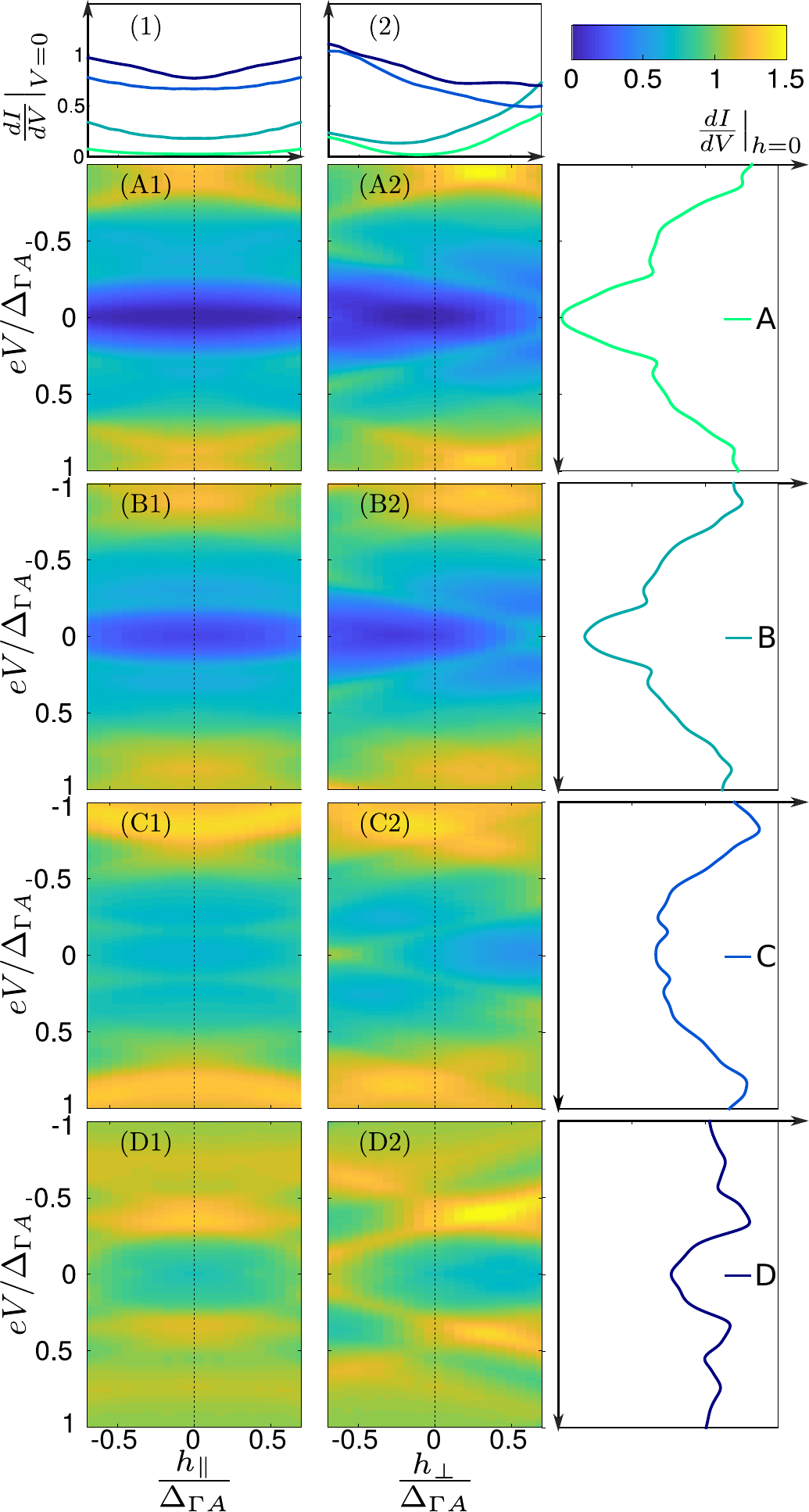}
\caption{Differential conductance for finite magnetic field for a model of Fe(Se,S)\cite{Setty2019} with parameter sets A-D: (1) conductance at zero energy as a function of in plane field normalized by the order parameter ($h_\parallel/\Delta_{\Gamma A}$)
(2) the same quantity for out of plane field, $h_\perp/\Delta_{\Gamma A}$. %
The slope at $h=0$ depends on the details of the order parameter. Other panels: false color maps of the conductance
for fields in plane $h_\parallel$ (A1-D1) exhibiting mirror symmetry at the dashed line and out of plane $h_\perp$ (A2-D2). Right panels: conductance as a function of bias voltage $V$, i.e. cut through the data at the dashed vertical lines in (A1-D1). (1) and (2) are cuts through the data at the horizontal line at $eV=0$.
} \label{fig_dos_field}
\end{figure}
The integration was performed from $-2\pi\le k_{x,y}\le 2\pi$. The upper and lower boundaries of the integral, $E_U, E_L$, were set to extend the plotted range over several scales of the temperature such that the derivative of the Fermi function outside that window is numerically zero. The calculation of LDOS was carried out on a momentum grid of size $800 \times 800$ points. The energy grid was spaced by $0.0015$ and artificial broadening $\eta$ was set to $0.0004$.

The results are presented in Fig. \ref{fig_dos_field}, where we first show the differential conductance (normalized to the normal state value) at zero voltage: Panel (1) shows this quantity as a function of magnetic field in plane for the four different parameter sets A-D that correspond
to different doping levels in our model for Fe(Se,S). In the ultranodal state, the conductance is nonzero, but is a symmetric function of the field $h_\parallel$. This is in contrast to the zero energy conductance for a field out of plane, which is not an even function of $h_\perp$. Note that the slope of the curves at $h_\perp=0$ strongly depends on the details of the model.
The false color maps show the dependence of the conductance as function of bias voltage and external field.
Panels (A1-D1) for in plane fields exhibit a mirror symmetry with respect to the dashed line, while this symmetry is absent in panels (A2-D2) for fields out of plane.
For convenience, we show the conductance at zero field in the right row which is a plot of the data from the false color maps along the dashed vertical line in each of the panels (A1-D2).
Note that qualitatively similar curves as in (1) and (2) are obtained for horizontal cuts at any bias $|V|\lesssim \Delta_{\Gamma A}$: The conductance at any bias voltage is an even function of the in plane field, but has odd powers for out of plane field,  a qualitatively different behavior than expected from low energy states due to vortices\cite{Volovik1993,CdGM1964}.

\subsection{Superfluid density}
\begin{figure*}[tb]
    \centering
    \includegraphics[width=\linewidth]{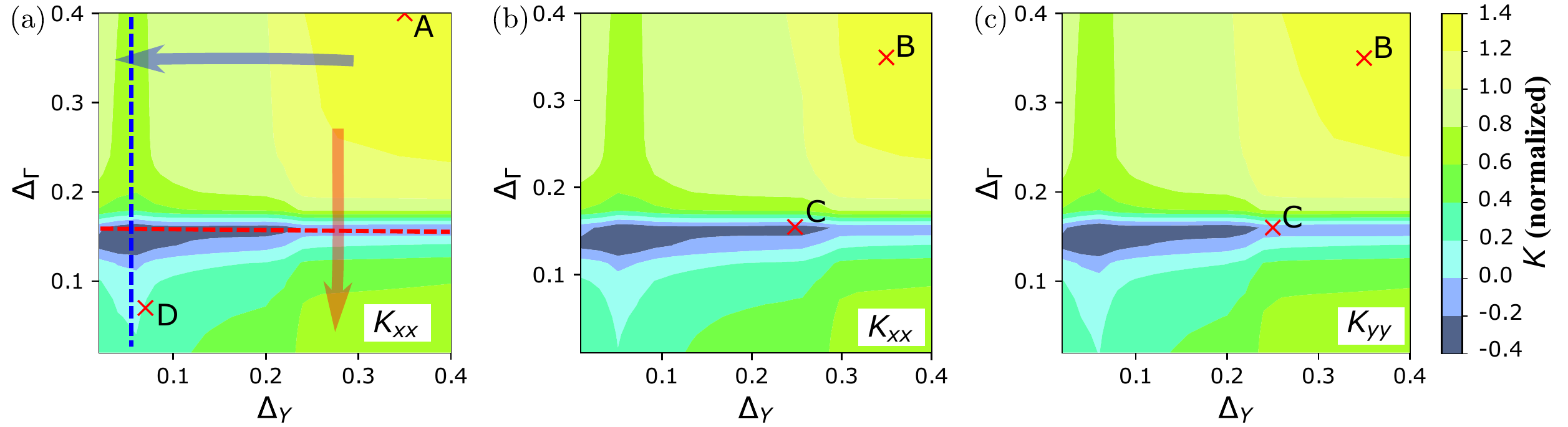}
    \caption{Contour plots of superfluid density as isotropic gap components vary. The inter-band gap component is $\Delta_0=0.4$ and the TRSB component $\delta=\Delta_0$.  Anisotropic gap components are $\Delta_{\Gamma a}=0.1, \Delta_{Xa}=\Delta_{Ya}=0.4$. The choice of parameters follows that of Fig. 2 in Ref.\onlinecite{Setty2019}, so that the shapes (or nonexistence) of Bogoliubov Fermi surfaces are known at specific points in this figure. (a) xx-component of the response kernel tensor $K_{xx}$, assuming $C_4$ symmetry is preserved. $\Delta_X=\Delta_Y$. (b)(c) $K_{xx}$ and $K_{yy}$, when $C_4$ symmetry is broken. $\Delta_X=0.8\Delta_Y$.}
    \label{sfdensity1}
\end{figure*}
In this section, we calculate the superfluid density to analyze the stability of the BFS states. BFSs found in our model can be unstable if they are associated with negative superfluid density. We show in this section that over a large parameter space, our model results in a positive superfluid density, and stable BFSs exist without the necessity of fine-tuning. On the other hand, zero temperature superfluid density is suppressed by the existence of BFSs since the quasiparticle excitations bear the same dimensionality of the normal state Fermi surface. This can in principle serve as an experimental signature.\par
We start with the current operator for a multi-band system with normal state dispersion $\epsilon_\alpha(\mathbf{k})$ in the presence of a vector potential with components $A_i$,

\begin{figure*}[tb]
\includegraphics[width=\linewidth]{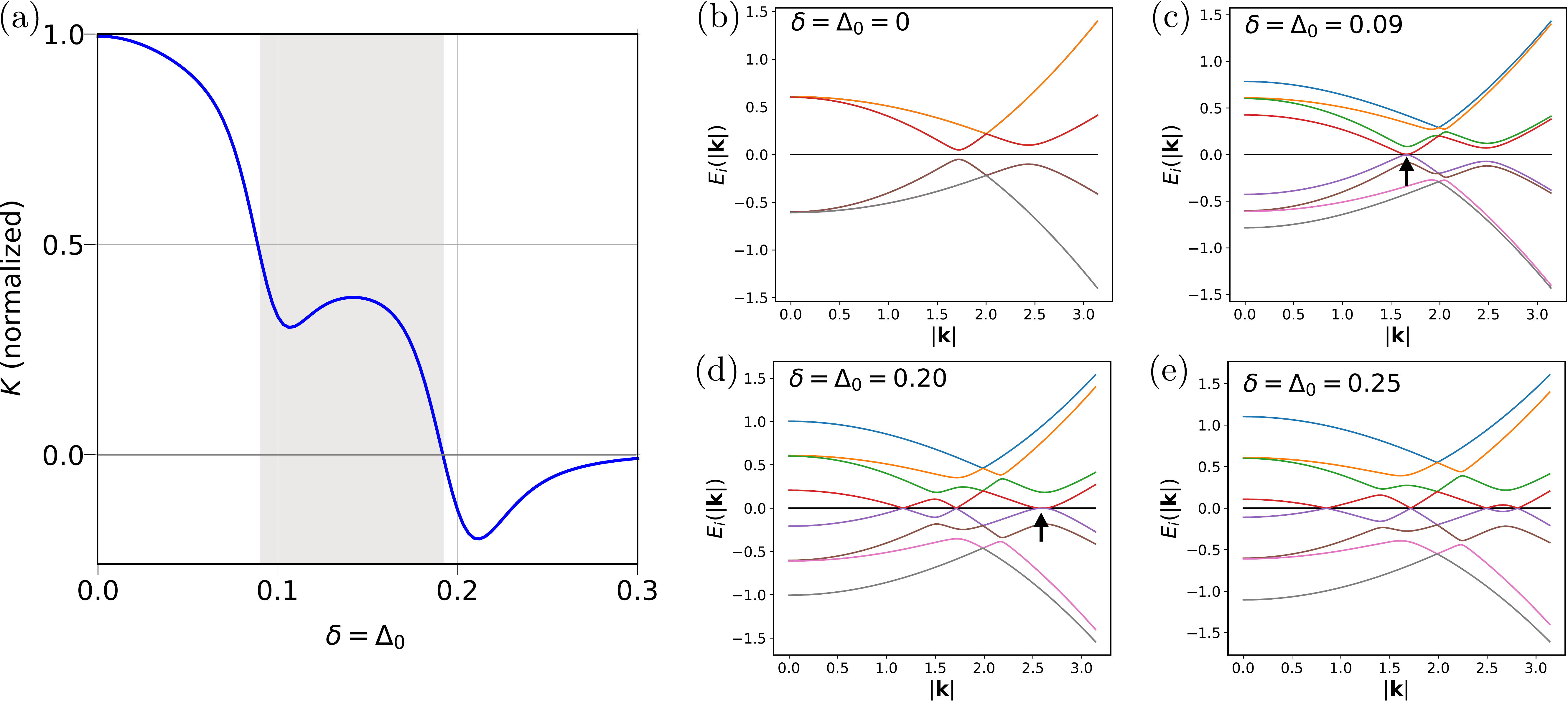}
    \caption{(a) Superfluid density of an isotropic two-band model.  The inter-band TRS and TRSB gap components $\Delta_0$ and $\delta$ are tuned and are set equal to each other. The intra-band gaps are $\Delta_1=0.05, \Delta_2=0.10$. Shaded area shows where stable BFS can exist with positive superfluid density. (b-e) Quasiparticle bands of the same isotropic two-band model plotted along radial $|\mathbf{k}|$ direction. Note the van Hove like singularity at $|\mathbf{k}|\approx1.6$ {\color{black} in (c)} , and that at $|\mathbf{k}|\approx2.7$ {\color{black} in (d)}, which give rise to the first and the second dip in (a) respectively.}
    \label{sfdensity2}
\end{figure*}

\beq
j^{p}_x(\mathbf{q}=0)=-\frac{e}{N_k}\sum_{\mathbf{k},\alpha,\sigma}\frac{\partial \epsilon_\alpha(\mathbf{k})}{\partial k_x}c^{\dagger}_{\mathbf{k}\alpha\sigma}c_{\mathbf{k}\alpha\sigma}\\
j^d_x(\mathbf{q}=0)=\frac{e}{cN_k}\sum_{\mathbf{k},\alpha,\sigma}\frac{\partial^2 \epsilon_\alpha(\mathbf{k})}{\partial k_x^2}A_x c^{\dagger}_{\mathbf{k}\alpha\sigma}c_{\mathbf{k}\alpha\sigma}
\eeq
$j^p, j^d$ are the paramagnetic and diamagnetic current respectively. The contributions to the current response kernel are accordingly
\begin{align}
K^p_{xx}(\mathbf{q}\rightarrow0,\omega=0)=\frac{\pi e^2}{c^2} \\
\times\frac{T}{N_k}\sum_{\mathbf{k}}\sum_{\nu_n}\Tr&(V_x(\mathbf{k})G(\mathbf{k},i\nu_n)V_x(\mathbf{k})G(\mathbf{k},i\nu_n))\,,\notag
\end{align}
where we defined a velocity matrix,
\beq
V_x(\mathbf{k})=\begin{pmatrix}
    \frac{\partial \epsilon_1(\mathbf{k})}{\partial k_x}\mathbb{1}& \\
     &\frac{\partial \epsilon_2(\mathbf{k})}{\partial k_x}\mathbb{1}& \\
     & & \ddots 
    \end{pmatrix}
\eeq
    and
\beq
    K^d_{xx}(\mathbf{q}\rightarrow0,\omega=0)=\frac{4\pi e^2}{c^2N_k}\sum_{\mathbf{k},\alpha}\abs{\frac{\partial^2 \epsilon_\alpha(\mathbf{k})}{\partial k_x^2}}n_\alpha(\mathbf{k})
\eeq
as it can be read off from the definition in linear response theory
\beq
j_i(\mathbf{q},\omega)=-\frac{c}{4\pi}K_{ij}(\mathbf{q},\omega)A_{j}(\mathbf{q},\omega)\\
j=j^p+j^d,\ \ \ \ \ \ \ K=K^p+K^d.
\eeq
Here $n_\alpha$ is the density of electrons (holes) in electronlike (holelike) band $\alpha$ in the normal state, $G(\mathbf{k},i\nu_n)$ is the Nambu Green's function. $K_{xy}=0$ for the quadratic dispersion used in our model. $K_{yy}=K_{xx}$ if $C_4$ symmetry is preserved. The response kernels are to be calculated at zero temperature, but  $T=0.001$ is chosen when performing the numerics. The numerically calculated response kernels are normalized with respect to the normal state diamagnetic kernel $K^d(\Delta=0)$, in other words,  superfluid densities are normalized to the normal state carrier density.\par
We see from Fig.~\ref{sfdensity1} that there exists a regime of inter-band pairing where the superfluid density remains positive when the system supports BFSs. \textcolor{black}{To be specific, the labels A and B mark parameters yielding a positive superfluid density with no or very small BFSs; marker D is a parameter set exhibiting positive superfluid density with significant BFSs around all the electron pockets and the center hole pocket, and finally, marker C is an example where BFSs exist, but  the system has a negative superfluid density. As a matter of fact, we find that the BFSs around the center hole pocket develop gradually as $\Delta_\Gamma$ decreases (red arrow in panel (a)), while almost independently, the BFSs around the four electron pockets grow if $\Delta_X$ and $\Delta_Y$ decrease (blue arrow in panel (a)). The superfluid density goes down and then back up, or we say it shows ``dips'' (indicated by the  red and blue dashed lines in panel (a)), in both of the above processes. This kind of behavior is better understood in a simplified isotropic two-band model (Fig. \ref{sfdensity2}). We find that the dips in superfluid density are common near where the BFS begins to show up in the parameter space. A careful analysis shows that the dips in the superfluid density are associated with 
``van Hove-like'' singularities in the quasiparticle bands (in contrast to a van Hove singularity in the usual sense, which is associated with a normal state band). Fig. \ref{sfdensity2} shows the van Hove-like singularities as indicated by the small arrows in panel (c) and (d), and how these van Hove-like singularities are associated with the dips in the superfluid density when $\delta=\Delta_0$ is tuned.}\par
 A possible interpretation of the region where superfluid density turns negative is that the system is unstable toward a phase where  spatially modulating gaps and BFSs coexist with TRSB. However, in principle, the system could get around such a phase by acquiring a  momentum-dependent inter-band order parameter $\Delta_0(\bs k)=\delta(\bs k)$. This can happen by avoidance of the van-Hove singularity at zero energy due to residual pairing~\cite{Moon2019} of Bogoliubov quasiparticles. Further investigations of this problem are underway.

\section{Finite bandwidth effects}
\begin{figure}[tb]
      \includegraphics[width=\linewidth]{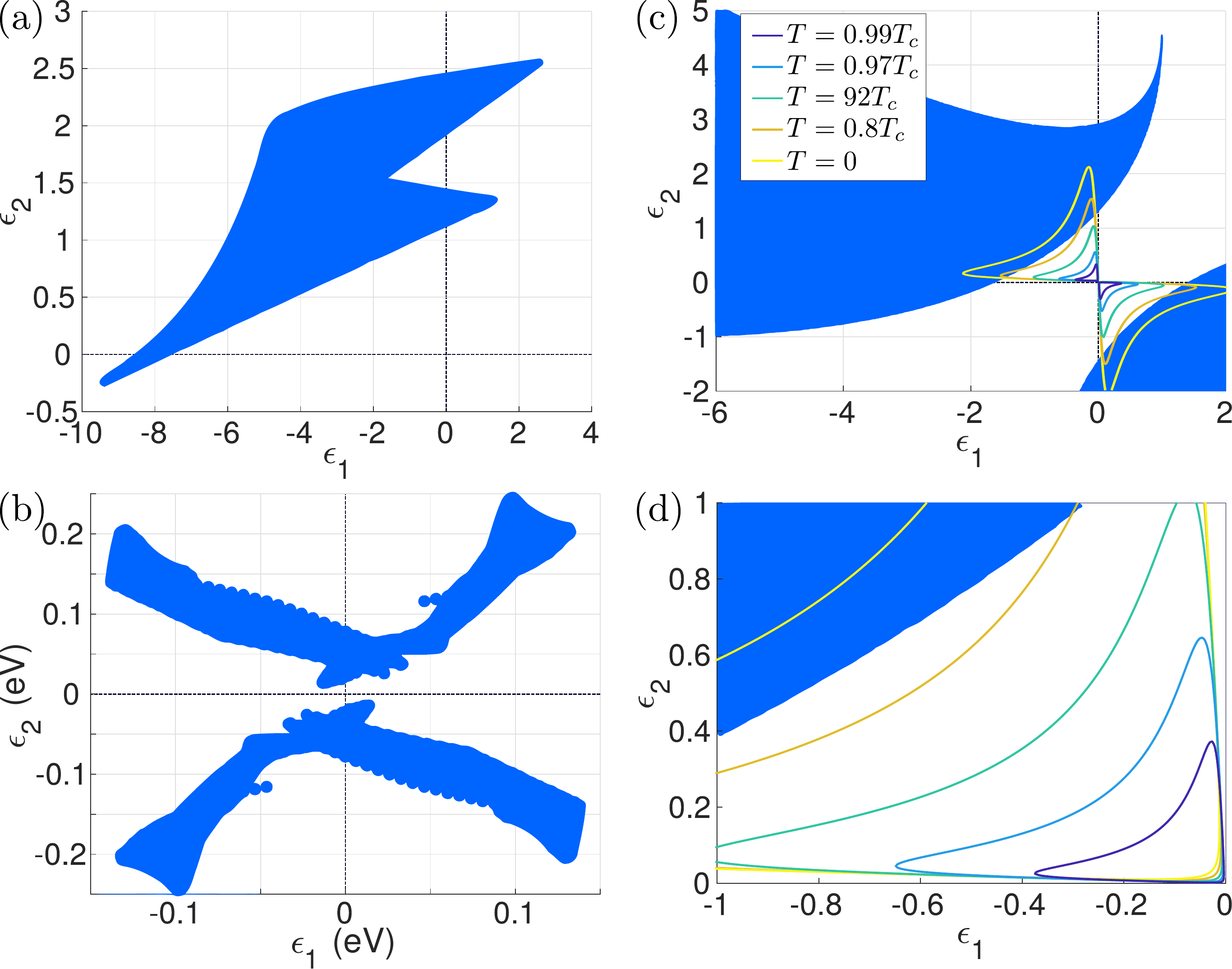}
  \caption{Transition to ultranodal state with finite bandwidth: Allowed lowest eigenenergies for electronic structures on a lattice for (a) two band model of Raghu et al.\cite{Raghu08} and (b) lowest two eigenenergies of a 5 band model for FeSe \cite{Sprau2017}. The blue area denotes the possible combinations of ($\epsilon_1$, $\epsilon_2$) in the respective electronic structure in the normal state. (c) Allowed energies of the model for Fe(Se,S)\cite{Setty2019} with electron and holelike bands together with contours where the Pfaffian is zero, i.e. inside the area of these contours the Pfaffian becomes negative. Assuming for simplicity isotropic intra-band order parameters $\Delta_1=\Delta_2=0.15$ and $\Delta=\delta=0.4$ at $T=0$ and inferring that all order parameters have the same (mean-field like) temperature dependence, the contours decrease in size and move towards the origin as temperature increases towards $T_{\mathrm c}$. Once, there is no allowed point ($\epsilon_1$, $\epsilon_2$) inside the contour, the system is not topological. (d) Blowup of the region close to the origin showing that the area of negative Pfaffian shrinks to a small triangle when $T\rightarrow T_{\mathrm c}$ and eventually does not any contain allowed ($\epsilon_1$, $\epsilon_2$).}
    \label{fig_allowed_e1_e2}
\end{figure}

In the calculation of the minimum of the Pfaffian as discussed in Sec. \ref{sec_zeeman}, the eigenenergies of the electronic structure were considered as free parameters and varied to find the minimum. For a real electronic structure, the band energies in the normal state of the two (or lowest two) bands are however not completely free parameters. Instead, there is a given relation $\epsilon_1(\mathbf{k})$ and $\epsilon_2(\mathbf{k})$ for all allowed momenta $\mathbf k$. Then, the minimization has to be done with the constraint on allowed band energies for the given model. In Fig. \ref{fig_allowed_e1_e2} the allowed pairs of ($\epsilon_1$, $\epsilon_2$) are shown as light blue area for representative models: (a) a minimal two band model for iron-based superconductors\cite{Raghu08}, (b) a realistic 5 band model for FeSe\cite{Sprau2017}, where only the lowest eigenvalues (in magnitude) have been considered and (c) the two pocket model of this work, Eqs. (\ref{eq_dispersion_two_pocket_1}-(\ref{eq_dispersion_two_pocket_3}). Note that not all area is covered and that there is a point reflection symmetry around the point (0,0) which simply reflects the interchange of $\epsilon_1$ and $\epsilon_2$. Considering now the model of the electronic structure in Eq. (\ref{eq_dispersion_two_pocket_1}-\ref{eq_dispersion_two_pocket_3}) and assuming that the superconducting order parameter exhibits the intra-band order parameters $\Delta_1$ and $\Delta_2$ together with a Type-1 TRSB, odd-parity, spin-triplet pair described by the order parameters $\Delta_0=\delta$, see Eq. (\ref{pairing_H})\cite{Setty2019}, one can calculate the contour lines in ($\epsilon_1$, $\epsilon_2$) where the Pfaffian vanishes. In Fig. \ref{fig_allowed_e1_e2} (c,d), the allowed pairs of eigenenergies represent the blue area and mentioned contour lines are plotted for a choice of $\Delta_1=\Delta_2=0.15$ and $\Delta_0=\delta=0.4$, which is the contour line at $T=0$. The area inside the contour lines exhibits $\mathrm{Pf}<0$, while outside $\mathrm{Pf}>0$. For $T>0$, we decrease all components of the superconducting order parameter according to the mean field behavior. Then, the size of the contour for $\mathrm{Pf}=0$ shrinks and the minimum of the Pfaffian moves along a trajectory $\epsilon_1=-\epsilon_2$ towards the origin, see Fig. \ref{fig_allowed_e1_e2} (d) such that at a temperature smaller than $T_{\mathrm c}$ there is no overlap between the allowed eigenenergies (light blue area) with the region where $\mathrm{Pf}<0$, i.e. the system enters the trivial state at $T<T_{\mathrm c}$. Note that unless there are two bands crossing the Fermi level at the same $\mathbf k$ value (accidental Fermi surface intersection), the transition to the ultranodal superconducting state is expected to happen at a temperature where the relevant superconducting order parameters are sizable. Thus we can conclude that closer the multiple bands are energetically, easier is the formation of BFSs. 

\section{Conclusions}
Despite several proposals for the realization of BFSs in materials,  conclusive evidence of their existence has been elusive. While energy and momentum resolved band structure probes such as ARPES will have the final say on this question, a consideration of combinations of indirect experimental probes that are sensitive to extended surfaces of gapless excitations in the superconducting state is urgent. This work attempts to make progress in this direction while also expanding on the class of model Hamiltonians which show topological transitions into the ultranodal state. Some of these Hamiltonians are commonly studied models in  the existing literature, while others include charge-conjugation or parity non-preserving terms that were not previously examined in the context of BFSs.  Here we have not commented on the microscopic method of generating these terms, but this is clearly an important further step to construct a convincing case for the existence of such states.

Our analysis of the effect of a weak Zeeman field on the electronic and thermodynamic properties of BFSs close to the topological critical point reveals a distinguishing feature of BFSs -- the dependence of residual observables such as the zero temperature specific heat or the zero frequency tunneling density of states on the sign of the out-of-plane external field. Generic features arising from the spin-resolved spectral functions can be verified using spin-polarized ARPES. 

 Our calculation of the total internal magnetisation in the ultranodal state shows that the non-unitary TRSB magnetic moment is small. This could make its experimental detection using standard probes such as $\mu$SR difficult. Our consideration of finite-band width effects on the Pfaffian sign-change condition identifies features of more realistic multiband models of Fe(Se,S) that support the BFSs as opposed to simplified two-band descriptions; additionally, multi-band systems where the bands are energetically closer to each other are more favorable to the formation of BFSs.

Finally, from our evaluation of the superfluid density in the ultranodal state, we can conclude that there exists a window of interband pairing strength for which BFSs are stable with positive phase stiffness. Outside this window, BFSs either do not exist or are unstable with negative superfluid density. Consequences of the latter to possible modulated superconductivity with broken time-reversal symmetry will be the subject of future work. In the meantime, more direct probes of BFSs such as ARPES and quantum oscillations could help paint a fuller picture of this rapidly developing story and pave the way toward a deeper understanding of the ultranodal state. \par
\textcolor{black}{\textit{Note Added:} After the completion of this paper, recent $\mu$SR measurements reported TRSB in sulfur doped FeSe~\cite{Shibauchi2020}. }\par
\textit{Acknowledgements -- } We are grateful for discussions with M. Sulangi, Y. Wang and K. Yang. CS, SB and PH are supported by the U.S. DOE grant number DE-FG02-05ER46236.
\bibliography{Bogoliubov.bib}

\begin{thebibliography}{33}
\bibitem [{Not()}]{Note}%
  \BibitemOpen
  \href@noop {} {\bibinfo  {journal} {From now on we only keep the lowest three
  eigenvalues in the effective model. We have checked that corrections from the
  high-energy bands are negligible.}\ }\BibitemShut {NoStop}%
\end{thebibliography}
\end{document}